\documentclass[useAMS,usenatbib]{mn2e_mod}
\usepackage{graphicx}
\usepackage{deluxetable}
\usepackage{threeparttable}
\usepackage{amsfonts,amsmath,amssymb}
\usepackage{epstopdf}

\newcommand{\memb}{M_{\rm emb}}
\newcommand{\mcore}{M_{\rm core}}
\newcommand{\sigz}{\Sigma_z}
\newcommand{\au}{{\rm AU}}
\newcommand{\gcr}{{\rm GCR}}
\newcommand{\rhog}{\rho_{\rm gas}}
\newcommand{\gcms}{{\rm g}/{\rm cm}^2}
\newcommand{\kep}{\emph{Kepler\ }}

\title[Rocky Recipe]{A Metallicity Recipe for Rocky Planets}
\author[Dawson, Chiang, \& Lee]{Rebekah I. Dawson$^{1}$\thanks{E-mail:
rdawson@berkeley.edu; Miller Fellow}, Eugene Chiang$^{1}$, and Eve J. Lee$^{1}$\\
$^{1}$University of California, Berkeley, 501 Campbell Hall \#3411, Berkeley, CA 94720-3411}
\date{Submitted to MNRAS on April 22, 2015. Accepted July 15,2015.}
\pagerange{\pageref{firstpage}--\pageref{lastpage}} \pubyear{2015}

\begin{document}
\maketitle
\label{firstpage}

\begin{abstract}
Planets with sizes between those of Earth and Neptune divide
into two populations: purely rocky bodies whose atmospheres
contribute negligibly to their sizes,
and larger gas-enveloped planets possessing
voluminous and optically thick atmospheres.
We show that whether a planet forms rocky or gas-enveloped
depends on the solid surface density of its parent disk.
Assembly times for rocky cores are
sensitive to disk solid surface density.
Lower surface densities spawn smaller planetary embryos;
to assemble a core of given mass, smaller embryos require
more mergers between bodies farther apart
and therefore exponentially longer formation times.
Gas accretion simulations yield a rule of thumb that a rocky core must be at least 2$M_\oplus$ before it can acquire a volumetrically significant atmosphere from its parent nebula.
In disks of low solid surface density, cores of such mass appear
only after the gas disk has dissipated, and so 
remain purely rocky. Higher surface density disks breed
massive cores more quickly, within the gas disk lifetime,
and so produce gas-enveloped planets. 
We test model predictions against observations,
using planet radius as an observational proxy for gas-to-rock content
and host star metallicity as a proxy for disk solid surface density.
Theory can explain the observation that metal-rich stars host
predominantly gas-enveloped planets.
\end{abstract}

\section{Introduction}

The \kep Mission has discovered an abundance of planets --- of order
one per star \citep{Fres13} --- with orbital periods
shorter than a year, and sizes smaller than that of Neptune
\citep{Boru11a,Boru11b,Bata13,Burk14,Mull15}. \kep measurements of
planet radii, when combined with planet
masses measured via radial-velocity follow-up
\citep{Marc14,Weis14,Dres15} and transit timing variations
\citep{Ford11,Fabr12,Ford12,Ford12a,Lith12,Stef12,Stef12a,Maze13,Stef13,Wu13,Hadd14,Xie14}, reveal a surprising
diversity of bulk densities. Some planets are dense enough to be made
practically exclusively of solid refractory elements --- e.g., Kepler-10b
\citep{Bata11} and Kepler-36b \citep{Cart12} --- whereas others require
voluminous volatile
components, e.g., the Kepler-11 planets \citep{Liss11,Liss13} and Kepler-36c
\citep{Cart12}. We refer to the former class as ``purely
rocky'' (a.k.a. ``super-Earths'')  and the latter as ``gas-enveloped'' (a.k.a.
``mini-Neptunes''). 

Purely rocky planets can undergo transformation to gas-enveloped, and vice
versa, through a variety of physical processes including outgassing
\citep{Roge11}, photoevaporation \citep{Lope12,Lope13}, and
collisional stripping of atmospheres \citep{Schl15}. Our interest here
is in determining a planet's make-up just after its formation.
We seek to identify the factors that dictate whether or not a rocky core can
accrete a volumetrically significant atmosphere from its parent nebula
\citep{Lee14}. We highlight the decisive role played by the surface
density of solids in the primordial disk --- and by extension
the host star metallicity --- in controlling how long
a rocky core takes to coagulate.
According to our interpretation, purely rocky planets are born within
disks of low solid surface density and accordingly have long gestation times:
they attain core masses
large enough to acquire atmospheres only after their ambient
gas disks have dissipated.
Recent breakthroughs linking planet radius to composition
\citep{Lope14,Roge15} allow us to identify with confidence
purely rocky vs.~gas-enveloped planets in the \kep sample for comparison
with theoretical models.

We present a
planet formation scenario in Section 2, deriving
approximate analytic estimates for the masses of the seed bodies or
``embryos'' that merge to form super-Earths
and the underlying cores of mini-Neptunes, and the ambient
gas densities during the merger era. These order-of-magnitude
considerations set the stage for numerical
simulations in Section 3, where we combine $N$-body integrations
with atmospheric accretion models to show that disks with the highest solid
surface densities spawn gas-enveloped planets, whereas disks having
sufficiently low solid surface densities produce purely rocky planets.
In Section 4, we translate, as best we can, model predictions 
for gas mass fraction vs.~disk surface density into the observables
of planet radius vs.~host star metallicity. There we 
compare to the observed sample of \kep candidates with spectroscopic host star metallicities \citep{Buch14} and find support for
the link between disk surface
density and 
small planets' composition.
We conclude in Section 5.

\section{Overview of formation scenario and initial conditions}
The solids in a protoplanetary disk provide the raw material for
building rocky planets and the cores of gas-enveloped planets. 
Planetesimals may have either formed in situ or accumulated
by radial drift from larger orbital distances.

In a disk of solid surface density $\sigz$,
an ``embryo'' (a.k.a.~isolation mass)
consumes all the solids within its feeding zone
of annular width $\Delta a$:
\begin{equation}
\memb = 2\pi a \Delta a \sigz.\label{eqn:memb}
\end{equation}
\noindent To give a sense of scale, if $\Delta a = 5 R_{\rm H}$, where
\begin{equation}
R_{\rm H} = a \left( \frac{2 \memb}{3 M_\star} \right)^{1/3}
\end{equation}
\noindent is the mutual Hill radius and $M_\star$ is the host star's
mass, then $\memb$ is approximately the mass of Mars at
semi-major axis $a =$ 0.3 AU in a disk with \mbox{$\Sigma_{z}=30\,\gcms\left( a/\au \right)^{-3/2}$} 
(for context, this surface density is about 3$\times$ that
of the solid component of the minimum-mass solar nebula;
see, e.g., \citealt{chia10} and use their $Z_{\rm rel} = 0.33$).
We will consider other
possibilities for $\Delta a$ and $\sigz$ below.

At short orbital periods, isolation-mass embryos grow quickly, consuming the planetesimals in their annuli well before the gas
disk dissipates. The growth or coagulation timescale is approximately
\begin{equation}
t_{\rm coag} \sim \frac{\rho_{\rm emb} R_{\rm emb}}{\sigz \mathcal{F} } \Omega^{-1},
\end{equation}
\noindent where $\Omega$ is the orbital frequency, $\rho_{\rm emb}$ the embryo's density, $R_{\rm emb}$ the embryo's radius, and $\mathcal{F} > 1$ is the enhancement of the growth rate due to gravitational focusing (e.g., \citealt{Gold04}). For example, even without gravitational focusing ($\mathcal{F} =1$),
the Mars-mass embryo mentioned above takes only 0.2 Myr to grow.

\subsection{Embryo self-stirring as the gas disk dissipates}
While the gas disk is still present, it damps the embryos'
eccentricities by dynamical friction, preventing them from crossing
orbits, merging, and growing to larger masses. The gas disk
needs to dissipate so that the embryos' self-stirring becomes more
effective against dynamical friction. The goal of this subsection is
to gauge the degree of gas depletion, and the magnitudes
of the embryos' random velocities and spacings, at the time that embryos
cross orbits --- i.e., at the onset of the ``giant impact'' stage of
planet formation. This will inform the choice of initial conditions
for the $N$-body coagulation simulations in Section 3 --- which will not include
gas dynamical friction. \footnote{Our analysis in 
Section 2 assumes that embryos do not 
open gaps in the gas disk. Whether this is true
depends on the disk viscosity (e.g., \citealt{Duff15},
and references therein). Gaps
weaken eccentricity damping by gas, but open
the possibility of eccentricity excitation by gas \citep{Gold03}. In any case,
we do not expect gaps to seriously interfere with
the ability of rocky cores to accrete gas envelopes,
since the gas accretion rate is insensitive
to the ambient gas density \citep{Lee14}
and since gas in the gap is continuously replenished
by viscous radial inflow \citep{Lubo06}.}

We sketch below
the various regimes for the embryos' equilibrium random velocities $v$.
There are three velocity scales.
The Hill velocity, $v_{\rm H}$, tends to be the smallest:
\begin{eqnarray}
v_{\rm H} & = & R_{\rm H} \Omega \nonumber \\
    & = & \left(\frac{2 \memb}{3M_\star}\right)^{1/3} a \Omega \nonumber \\
    &= &0.4 \,{\rm km/s} \,\left(\frac{\memb}{M_\oplus}\right)^{1/3} \left(\frac{M_\star}{M_\odot}\right)^{1/6} \left(\frac{\au}{a}\right)^{1/2} \,. \nonumber \\
\end{eqnarray}
The sound speed, $c_{\rm s}$, is not much larger.
We assume a disk temperature $T$ of
\begin{equation}
T = 1500 \, {\rm K} \, \left[\frac{0.1 \,\au}{ \max (a,0.1\au) }\right]^{1/2}, \label{eqn:temp}
\end{equation}
\noindent yielding a sound speed
\begin{equation}
c_{\rm s} = \sqrt{kT/\mu} = 1.29 \, {\rm km/s} \, \left[\frac{1 \, \au}{ \max(a, 0.1\au) }\right]^{1/4}, \label{eqn:cs}
\end{equation}
where $k$ is the Boltzmann constant and
\mbox{$\mu=2.34\times1.67\times10^{-24}$ g} is the mean molecular mass.

The largest scale is the escape velocity from the surface of the embryo:
\begin{eqnarray}
v_{\rm esc} &=& \sqrt{ \frac{2 G M_{\rm emb}}{R_{\rm emb}} }  \nonumber \\
 &= &11 \, {\rm km/s}\, \left(\frac{\memb}{M_\oplus}\right)^{1/2} \left(\frac{R_{\rm emb}}{R_\oplus}\right)^{-1/2}. \nonumber \\
\label{eqn:vesc}
\end{eqnarray}

\subsubsection{Growth of random velocities to $v_{\rm H}$}
During the first stage of stirring, the embryos' random velocities are below $v_{\rm H}$. Their stirring rate is 
\begin{equation}
\left(\frac{\dot{v}}{v}\right)_{\rm stir, 1} \sim \frac{\sigz}{\rho_{\rm emb} R_{\rm emb}} \frac{v_{\rm esc}}{v} \left(\frac{v_{\rm esc}}{v_{\rm H}}\right)^3 \Omega , v < v_{\rm H}
\end{equation}
\citep{Gold04}.
At the same time, gas damps random velocities at a rate 
\begin{equation}
\left(\frac{\dot{v}}{v}\right)_{\rm damp, 1} \sim - \frac{\Sigma_{\rm gas}}{\rho_{\rm emb} R_{\rm emb}} \left(\frac{v_{\rm esc}}{c_{\rm s}} \right)^4 \Omega , v < c_{\rm s} \label{eqn:gdamp}
\end{equation}
\citep[e.g.,][]{Komi02},
resulting in an equilibrium,
$$v_{\rm equi, 1} \sim \frac{\sigz}{\Sigma_{\rm gas}} \left(\frac{c_{\rm s}}{v_{\rm H}}\right)^3 c_{\rm s}, v < v_{\rm H}.$$
The embryos' random velocities grow to $v_{\rm H}$ once the gas surface density drops to
$$\Sigma_{\rm gas} \lesssim \sigz \left(\frac{c_{\rm s}}{v_{\rm H}}\right)^4 \,.$$
\noindent This condition evaluates to $\Sigma_{\rm gas} \lesssim 700
\sigz$ for Mars-mass embryos at $a=0.3\,\au$,
or $\Sigma_{\rm gas} \lesssim
100 \sigz$ for Earth-mass embryos at 1 AU.
For disks initialized with cosmic gas-to-solid ratios of $\Sigma_{\rm
  gas}/\sigz \sim 200$, little or no depletion is required for $v \sim v_{\rm H}$.

\subsubsection{Growth of random velocities to $c_{\rm s}$} \label{grow_cs}
Next the big bodies stir each other from random velocities of $v_{\rm H}$ to $c_{\rm s}$. The gas damping rate remains the same (Equation \ref{eqn:gdamp}), but the stirring rate decreases to 
\begin{equation}
\label{eqn:final_stir}
\left(\frac{\dot{v}}{v}\right)_{\rm stir,2} \sim  \frac{\sigz}{\rho_{\rm emb} R_{\rm emb}} \left(\frac{v_{\rm esc}}{v}\right)^4 \Omega, v_{\rm H} < v < v_{\rm esc}
\end{equation}
\citep{Gold04}, resulting in a new equilibrium
$$v_{\rm equi,2} \sim \left(\frac{\sigz}{\Sigma_{\rm gas}}\right)^{1/4} c_{\rm s}, v_{\rm H} < v < c_{\rm s}.$$
Random velocities reach $v \sim c_{\rm s}$ once $\Sigma_{\rm gas} \lesssim \sigz$.

\subsubsection{Growth of random velocities to $\sqrt{3}c_{\rm s}$} \label{grow_sqrt3}
Once the gas density drops enough for $v$ to exceed $c_{\rm s}$, the gas
damping rate decreases to the classical formula for dynamical friction
\citep{Papa00}. We assume the random energy is
equipartitioned so that the random vertical velocity is $c_{\rm s}/\sqrt{3}$
and the embryo remains marginally embedded in gas. The new damping
rate is
\begin{eqnarray}
\left(\frac{\dot{v}}{v}\right)_{\rm damp,3} \sim - \frac{\Sigma_{\rm gas}}{\rho_{\rm emb} R_{\rm emb}} \left( \frac{v_{\rm esc}}{c_{\rm s}} \right) \left( \frac{v_{\rm esc}}{v} \right)^3 \Omega \,, \nonumber \\
c_{\rm s} < v < \sqrt{3} c_{\rm s} , \nonumber \\
\end{eqnarray}
and the random velocity reaches a new equilibrium of 
$$v_{\rm equi,3}  \sim \frac{\sigz}{\Sigma_{\rm gas}} c_{\rm s}, c_{\rm s} < v < \sqrt{3} c_{\rm s}$$
until $\Sigma_{\rm gas} \sim \sigz/\sqrt{3}$.

\subsubsection{Gas damping shuts off} \label{shutoff}
Finally, once $\Sigma_{\rm gas} \lesssim \sigz / \sqrt{3}$ so that
$v \gtrsim \sqrt{3} c_{\rm s}$, the embryo is no longer embedded in the disk but plunges through the disk twice per orbit, reducing the gas damping further to 
\begin{equation}
\left(\frac{\dot{v}}{v}\right)_{\rm damp,4}  \sim - \frac{\Sigma_{\rm gas}}{\rho_{\rm emb} R_{\rm emb}}
\left(\frac{v_{\rm esc}}{v} \right)^4 \Omega ,  \sqrt{3} c_{\rm s} < v
\end{equation}
\citep{Ford07,Rein12}. At this stage, stirring (Equation \ref{eqn:final_stir}) always exceeds gas dynamical friction; gas damping effectively shuts off and 
cannot establish a velocity equilibrium.

\subsubsection{The case of widely spaced embryos}

The stirring rates cited above (Equations 8 and 10)
are based on the particle-in-a-box approximation, which
holds only for embryos spaced closely enough \citep{Ford07}. Spacings widen as embryos merge, eventually
invalidating the approximation. We estimate the reduced
stirring rate by using the encounter map of \citet{Hase90}.
Two planets separated in semimajor axis by $n R_{\rm H}$
undergo close encounters every synodic period
$\Delta t_{\rm s} = 4\pi a/ (3 n R_{\rm H} \Omega)$.
Each encounter increases planetary eccentricities by
$\Delta e$ as given by
\begin{equation} \label{eqn:hase90}
\Delta (ae/R_{\rm H})^2 \approx 2(a/R_{\rm H})^2 e \Delta e \approx \frac{45}{n^4} 
\end{equation}
\noindent (\citealt{Hase90}, their equation 38).
The stirring rate is then
\begin{equation}
\frac{\dot{v}}{v} \sim \frac{1}{e} \frac{\Delta e}{t_{\rm s}} \sim \frac{135 R_{\rm H}^3 \Omega }{8\pi n^3 e^2 a^3} \,. 
\end{equation}
\noindent Substituting
\begin{equation}
\Sigma_z = \frac{2R_{\rm emb}^3 \rho_{\rm emb}}{3 a n R_{\rm H}}
\end{equation}
\noindent yields
\begin{equation}
\label{eqn:vstir5}
\left(\frac{\dot{v}}{v}\right)_{\rm stir, 5} \sim  \frac{\Sigma_z}{\rho_{\rm emb} R_{\rm emb}}\, \frac{v_{\rm esc}^4}{v^2 n^2 v_{\rm H}^2} \,\Omega  \,. 
\end{equation}
This formula is probably an upper limit to the true
stirring rate, as it assumes that the planets'
eccentricity and inclination vectors are randomized
between encounters (i.e., Equation \ref{eqn:hase90}
is ``phase-averaged").

For $v<c_s$, equating Equation \ref{eqn:vstir5} to the damping rate in Equation 9 results in a velocity equilibrium
\begin{equation}
v_{\rm equi, 5} = \sqrt{\frac{\Sigma_z}{\Sigma_{\rm gas}}} \frac{c_{\rm s}^2}{ n v_{\rm H}} \,. 
\end{equation}
\noindent The equilibrium velocity reaches $c_{\rm s}$
when the gas surface density drops to 
\begin{equation}
\label{eqn:ratwide}
 \frac{\Sigma_{\rm gas}}{\Sigma_z} <  \left( \frac{c_{\rm s}}{n v_{\rm H}} \right)^2  \,. 
\end{equation}
\noindent This condition evaluates to $\Sigma_{\rm gas} \lesssim \Sigma_z$ for Mars-mass embryos at 0.3 AU
separated by $n=5$.

Once the equilibrium velocity exceeds
$c_{\rm s}$, the ratio of the stirring rate (Equation 16) to the damping
rate (Equation 11) grows with $v$, and
the velocity equilibrium is no longer stable.
In this case, gas damping shuts off. However,
it may not stay shut off.
Equation \ref{eqn:ratwide} suggests that as embryos grow more
widely spaced through successive mergers, 
eccentricity damping by gas may return to
significance, bringing the equilibrium velocity
back below $c_{\rm s}$ and necessitating further
reductions in $\Sigma_{\rm gas}$ to effect
more mergers. For example, Equation \ref{eqn:ratwide} evaluates to $\Sigma_{\rm gas} \lesssim 0.04 \Sigma_z$ for $2 M_\oplus$
cores separated by 10 $R_{\rm H}$ at 0.3 AU.
Intermittent gas damping would both increase the total core growth timescale and limit the amount of gas a planet can ultimately accrete. We will not account for such ongoing gas damping in the
remainder of the paper, but a detailed, self-consistent treatment for parameters appropriate to close-in super-Earths is an important subject for future work (see, e.g., \citealt{Komi02} for a study pertinent
to solar system terrestrial planets).

\subsection{Growth to super-Earths and mini-Neptunes in the giant impact stage}
The results of the previous section suggest that, as a rough
rule of thumb,
$\Sigma_{\rm gas} \sim \sigz$ is the condition that
deactivates gas dynamical friction and that
triggers when embryos can cross
orbits and merge; furthermore, at the start of this giant
impact stage, random velocities $v \sim c_{\rm s}$. An alternative scaling for the random velocity is to use
$v_{\rm H}$;
\citet{Ida93} find that Hill scalings are appropriate for
mutual
stirring of oligarchs in the presence of planetesimals
(neglecting gas). We will experiment with both scalings,
respectively, in sections \ref{init_hill} and \ref{init_cs}.

With $\Sigma_{\rm gas} \sim \sigz$, the giant impact stage
commences with ample gas for merging embryos to
accrete. How much gas the growing rocky cores actually accrete
depends on how long the merging process takes.
If a super-Earth-mass core takes too long to coagulate, the already depleted gaseous nebula will
have completely dissipated (by viscous accretion onto the star) before
the core can acquire a volumetrically significant atmosphere.

The timescale $T$ for embryos to cross orbits is \citep{Yosh99}:
\begin{equation}
\log_{10} T = C_1 + C_2 n
\end{equation}
where $C_1$ and $C_2$ are constants that depend on the initial random velocities --- in units of $v_{\rm H}$ --- and $n$ is the number of mutual Hill radii $R_{\rm H}$ between embryos. Therefore for a given set of embryo separations and initial velocities scaled by their Hill values, the timescale for the first merger is independent of $\sigz$. But for the small values of $\memb$ in low $\sigz$ disks, one merger is not enough to build super-Earth cores; the timescale to grow to $\mcore > \memb$ depends strongly on $\sigz$, as we now show.

\subsubsection{Initial conditions set by the Hill scale} \label{init_hill}
We start by considering the case where the initial $C_1$, $C_2$, and $n$ are independent of $\sigz$. For example, \citet{Hans12} and \citet{Hans13} recently simulated the giant impact stage with initial orbital spacings defined using a fixed multiple of $R_{\rm H}$, and random velocities independent of $\sigz$. In this set up, the first orbit crossing time $T$ is independent of $\sigz$. However, the number of mergers will depend on $\sigz$. For an initial spacing of $n_0$ Hill radii, the embryo mass (Equation \ref{eqn:memb}) is 
\begin{eqnarray}
\memb =
0.16 M_{\oplus} \left(\frac{n_0}{10}\right)^{3/2}\left(\frac{\Sigma_{z,1}}{10\, {\rm g}/{\rm cm}^2}\right)^{3/2} \nonumber \\
\times \left( \frac{a}{\au}\right)^{\frac{3}{2}(2-\alpha)} \left( \frac{M_\star}{M_\odot} \right)^{-1/2} 
\end{eqnarray}
where $\sigz = \Sigma_{z,1} (a/{\rm AU})^{-\alpha}$ \citep[e.g.,][]{Koku02}.

To grow to $\mcore$ --- here loosely defined as the minimum mass necessary to accrete a volumetrically significant atmosphere --- equal-mass embryos must undergo approximately $\log_2(\mcore/\memb)$ mergers, each of which increases the number of Hill spacings $n$ by $2^{2/3}$ (since each merger increases the absolute separation between embryos by a factor of 2 while the mutual Hill radius increases by a factor of $2^{1/3}$). The timescale for the final merger to form $\mcore$ is therefore
$$\log_{10} T_{\rm final} = C_1 + C_2 \left(\mcore/\memb\right)^{2/3} n_0,$$
which we can rewrite as
$$\log_{10} T_{\rm final} = C_1 + C_3/\sigz,$$
where $C_3$ is independent of $\sigz$. Therefore the time to form $\mcore$ depends exponentially on the solid surface density.

\subsubsection{Initial conditions set by gas sound speed}\label{init_cs}
If the initial spacings and random velocities are scaled instead by the
gas sound speed $c_{\rm s}$ (sections \ref{grow_cs}--\ref{shutoff}), then
$$\memb \sim 2 \pi a \sigz c_{\rm s}/\Omega$$
and
$$n_0 = \frac{c_{\rm s}}{\Omega R_{\rm H}} \,.$$
The timescale for the final merger is
$$\log_{10} T_{\rm final} \sim C_1 + C_2 \frac{\mcore^{2/3} M_\odot^{1/3} }{\memb}\frac{c_{\rm s}}{\Omega a}\,.$$
The coefficient $C_2$ may also depend on $\sigz$, because $c_{\rm s}$ does not scale with $v_{\rm H}$. \citet{Yosh99} found that $C_2 \approx 0.84 (1-0.8 \langle \tilde{e}^2\rangle_{\rm initial}^{1/2})$, where $\tilde{e} = e a/R_{\rm H}$. Then for $\tilde{e} \sim c_{\rm s}/v_{\rm H}$,
$$\log_{10} T_{\rm final} = C_1 + \frac{C_3}{\sigz}\left[1-\frac{C_4}{\sigz^{1/3}}\right]$$
where $C_3$ and $C_4$ are independent of $\sigz$. Alternatively, if initial random velocities are damped to zero by mergers that precede the final doubling, then
$$\log_{10}  T_{\rm final}= C_1 + C_3/\sigz\,.$$
Either way, we see that the final merger to reach $M_{\rm core}$ takes a time that is exponentially
sensitive to $\sigz$.

\section{Simulations of core growth and gas accretion}
We saw in Section 2 that embryos begin crossing orbits
and merging when the gas surface density $\Sigma_{\rm gas}$
drops below the solid surface density $\sigz$. The subsequent
merger history unfolds over a timescale that is exponentially
sensitive to $\sigz$. High $\sigz$ disks promise to spawn
massive cores quickly enough that they can accrete volumetrically
significant atmospheres before the gas disk dissipates.
Here we determine using $N$-body coagulation simulations
whether this promise can be fulfilled.

\subsection{$N$-body simulations of core growth}
\begin{figure}
\begin{center}
\includegraphics{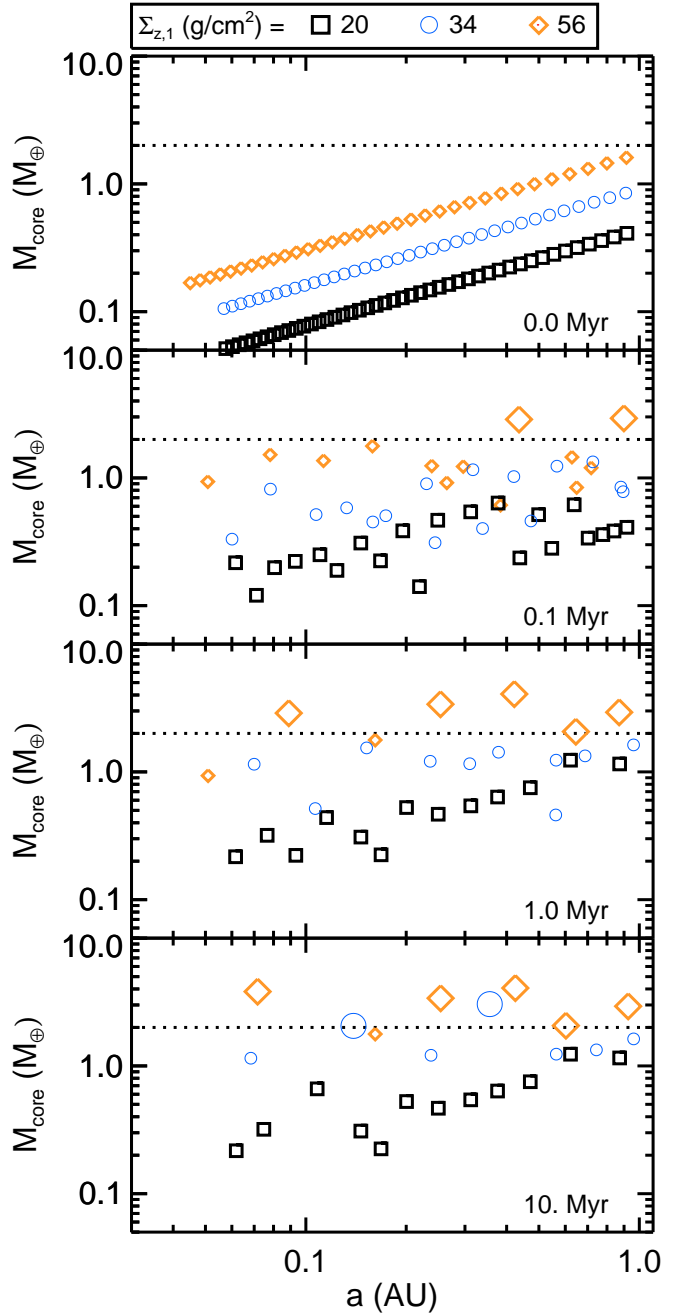}
\caption{\label{fig:ex} Snapshots of coagulating protocore masses vs.~semi-major axis from three simulations, color-coded by solid surface density normalization $\Sigma_{z,1}$. For reference, the minimum-mass solar nebula has $\Sigma_{z,1} \simeq 11$ g/cm$^2$.
Snapshot times are listed vertically on the right. Planets in the high solid surface density simulation (orange diamonds) grow quickly and reach $2 M_\oplus$ (dotted horizontal line) well within the gas disk lifetime of $\sim$1 Myr, matriculating to become gas-enveloped mini-Neptunes
(large symbols). For the intermediate solid surface density (blue circles), planets grow to $2 M_\oplus$ only after millions of years. In the lowest solid surface density disk (black squares), planets fail to reach $2 M_\oplus$ over the 27 Myr duration of the simulation. }
\end{center}
\end{figure}
We run four sets of $N$-body integrations using {\tt mercury6} \citep{Cham99} to simulate the era of giant impacts, similar to those run by \citet{Cham98} and \citet{Hans12,Hans13}. We use the hybrid symplectic integrator with a time step of 0.5 days and a close encounter distance (which triggers a transition from the symplectic integrator to the Burlisch-Stoer integrator) of 1 $R_{\rm H}$. In each set of 500 integrations, we begin with isolation-mass embryos resulting from disk solid surface densities spanning 3--400 g/cm$^2$ at 1 AU ($\equiv \Sigma_{z,1}$). All surface density profiles scale as $a^{-3/2}$.

In the first set of simulations (corresponding to the scaling arguments in Section 2.2.1), the embryo masses are defined by annuli each spanning 10 mutual Hill radii $(10 R_{\rm H})$, similar to \citet{Hans12,Hans13} and \citet{Koku98,Koku02}. The embryos' initial semi-major axes span $0.04 < a < 1 \,\au$. Initial random velocities are set to the Hill velocity $v_{\rm H}$, equipartitioned between eccentricity ($e = \sqrt{2/3}v_{\rm H}$) and inclination ($i = v_{\rm H}/\sqrt{3}$). In the second set of simulations (corresponding to Section 2.2.2), the embryo masses are defined by annuli spanning $2 c_{\rm s}/\Omega$, and random velocities are set to $c_{\rm s}$ (Equation \ref{eqn:cs}), similarly equipartitioned. 
In these first and second sets of simulations, all bodies have bulk
densities set to 1 g/cm$^3$, which defines their collisional
cross-sections. The third and fourth set of simulations 
correspond to the first and second sets, respectively, except with bulk
densities of 5 g/cm$^3$ instead of 1 g/cm$^3$. (For comparison, \citet{Hans13} use 3 g/cm$^3$.) Because the results did not change
materially upon using the higher bulk density, all of the results
shown below will be drawn from our first and second simulation sets.

\begin{figure*}
\begin{center}
\includegraphics[width=2\columnwidth]{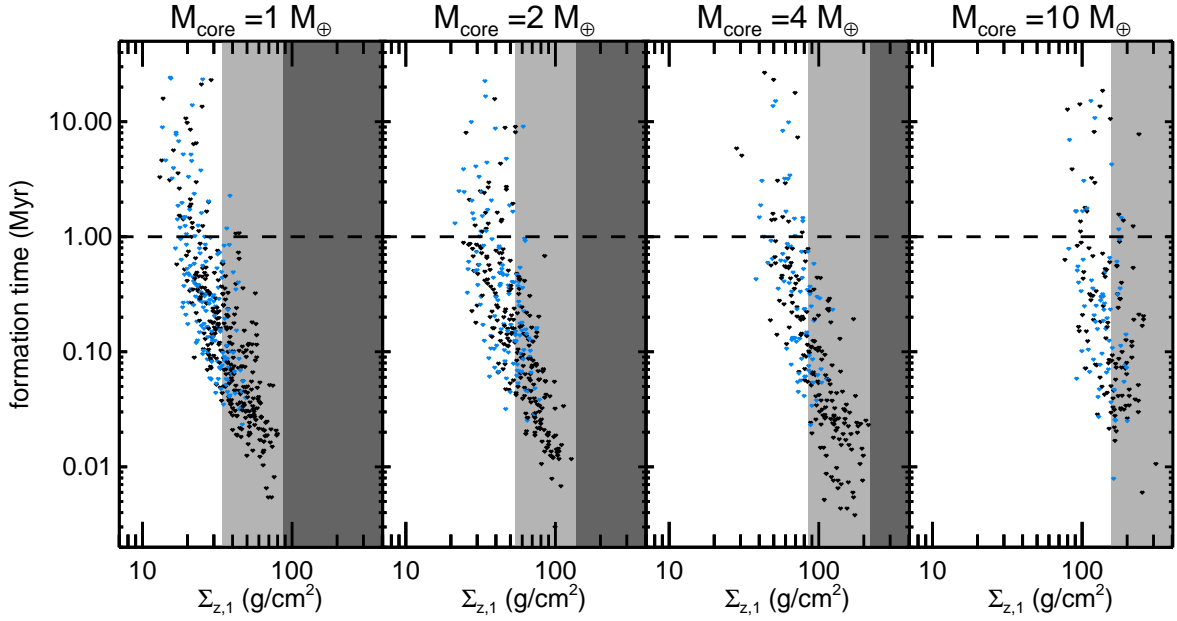}
\caption{\label{fig:growh} Time to grow to $\mcore$ (as listed on the top of the figure) as a function of solid surface density normalization $\Sigma_{z,1}$. The results shown here are computed from $N$-body integrations with embryo masses defined by annuli 10$R_{\rm H}$ wide, initialized with random velocities $v_{\rm H}$. The dashed horizontal line marks 1 Myr, an approximate remaining lifetime for the gas disk. The light gray region demarcates those surface densities for which $\memb=\mcore$ at $t=0$ (i.e., no mergers required to grow to $\mcore$) from $a=1\,\au$ (left boundary) to $a=0.15\,\au$ (right boundary). In the dark gray region, all embryos at all semi-major axes
start with $\memb\ge\mcore$ and therefore no points are plotted. Blue points have $0.5 < a(\au) < 1$ and black points have $0.15 < a(\au) < 0.5$. In the lower solid surface density disks, formation timescales are sometimes much longer than 1 Myr (top left of each panel).
}
\end{center}
\end{figure*}

\begin{figure*}
\begin{center}
\includegraphics[width=2\columnwidth]{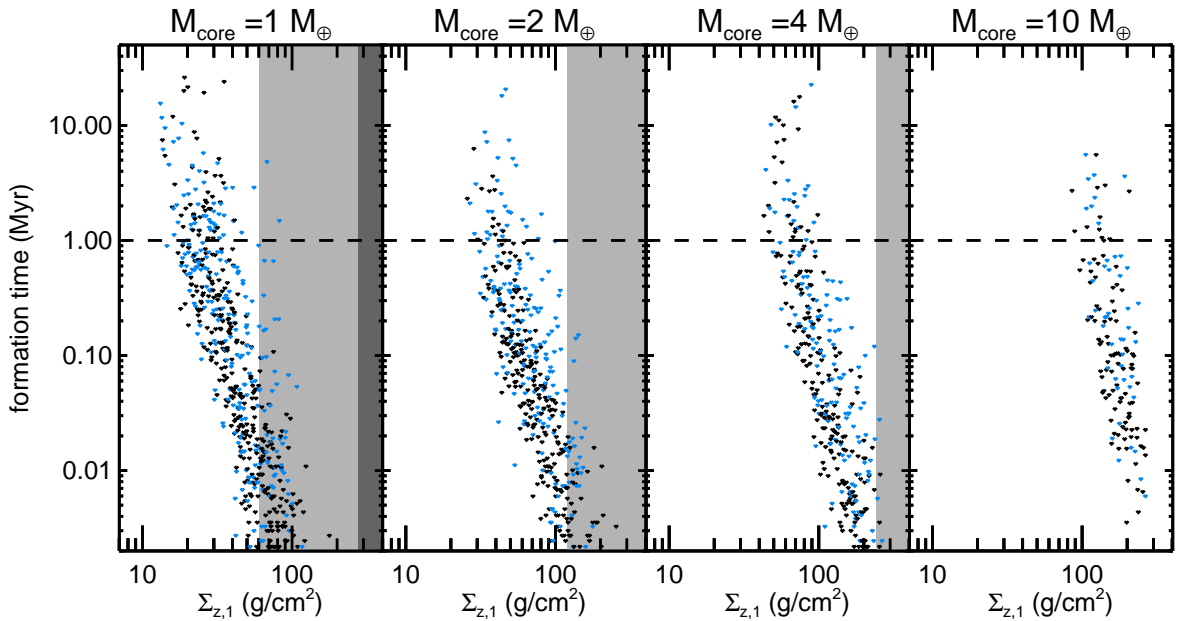}
\caption{\label{fig:growc} Same as Figure \ref{fig:growh} but for embryo
masses defined in annuli $2c_{\rm s}/\Omega$ wide, initialized with
random velocities $c_{\rm s}$. 
The growth timescales exhibit qualitatively the same features as in Figure
\ref{fig:growh} except they depend less sensitively on semi-major axis (the
black and blue points nearly overlap); embryo masses defined using $c_{\rm s}$ do not
vary as strongly with semi-major axis.}
\end{center}
\end{figure*}

The solid surface density, by means of setting the initial embryo masses, strongly affects the core coagulation timescale. Smaller
initial embryos necessitate more mergers to reach a given core mass. More mergers take longer time (exponentially longer, according
to the rough arguments presented in Section 2). In Figure \ref{fig:ex}, we show example snapshots from three simulations (all with Hill-scaled initial conditions) of disks with different $\Sigma_z$. Cores quickly reach substantial masses $(M_{\rm core} > 2 M_\oplus)$ in the highest $\Sigma_z$ disk; take millions of years to form in a lower $\Sigma_z$ disk; and remain low mass in the lowest $\Sigma_z$ disk. Next we summarize the results of two sets of 500 simulations a piece. In Figures \ref{fig:growh} and \ref{fig:growc}, we plot the time to grow to $\mcore$, as a function of $\Sigma_{z,1}$, for four sample values of $\mcore$. Figure \ref{fig:growh} corresponds to Hill-scaled initial conditions and Figure \ref{fig:growc} to those set by $c_{\rm s}$. A horizontal dashed line marks an approximate remaining lifetime of 1 Myr for the gas disk once $\Sigma_{\rm gas}$ has declined to $\sigz$. We employ only those simulations for which the maximum core mass attained is less than 30 Earth masses. For a wide range of $\sigz$, the embryos can grow to $\mcore$ in sufficiently short time to accrete atmospheres and become enveloped by gas; in some cases, the embryo mass itself exceeds $\mcore$ (gray regions). But for a subset of lower $\sigz$ disks, the time to grow to $\mcore$ is much longer than the remaining gas disk lifetime; such planets run out of time to accrete gas and remain purely rocky.

\subsection{Atmospheric accretion models}\label{aam}

\citet{Lee14} computed how a rocky core of a given mass
accretes gas from the primordial nebula, with specific application to \kep planets. We use their
model to generate, for an array of core masses, a set of gas accretion histories that we will combine in Section 3.3
with the $N$-body coagulation simulations from Section 3.1. This combination will enable us to determine, for
any given protoplanet,
the simultaneous time evolution of core mass $M_{\rm core}(t)$
and atmospheric mass fraction GCR$(t) \equiv$ gas-to-core
mass ratio.

For each core mass $\mcore \in \{1, 2, 4, 10\} M_\oplus$, we run
the \citet{Lee14} model for two nebular temperatures
$T \in \{ 900, 600 \}$ K appropriate for two orbital
distances $a \in \{0.3, 0.7\} \au$, respectively.
For every $T$ and $\mcore$, we adopt
two nebular gas densities. The higher gas density $\rhog = \Sigma_{\rm gas}\Omega/c_{\rm s}$ is chosen such that
$\Sigma_{\rm gas} = \sigz$, where $\sigz$ is that (extreme)
value required to form $\mcore$ as an isolation-mass embryo
from a $10$-$R_{\rm H}$-wide annulus at the given $a$.
The lower gas density is $1/100$ this value.
The set of models so constructed
span a large enough range of $\rhog$ and $M_{\rm core}$
that they can be usefully interpolated. 
We run one set of $4 [M_{\rm core}] \times 2 [T(a)] \times 2 [\rhog] = 16$ gas
accretion
simulations using solar metallicity opacities with
dust, and another set of 16 simulations using solar metallicity
opacities without dust (for technical details,
see \citealt{Lee14} and \citealt{Ferg05}).
Model parameters are summarized in Table \ref{tab:accrete}, including the corresponding values of $\Sigma_{z,1}$.

\begin{table}
\caption{Parameters for models of gas accretion in Figure \ref{fig:atm} \label{tab:accrete}}
\begin{threeparttable}[b]
\begin{tabular}{@{}lllll@{}}
\hline
$a$ (AU) & $T$ (K) & $\mcore (M_\oplus)$ & $\rhog$ ($10^{-12}$ g/cm$^3$) & $\Sigma_{z,1} ({\rm g}/{\rm cm}^2) $\\
\hline
0.7 & 600 & 1 & 65 &41 \\
&&&0.65&\\
&&2&100&64\\
&&&1.0&\\
&&4&160&102\\
&&&1.6&\\
&&10&300&188\\
&&&3&\\
0.3 & 900 & 1 &1000&62\\
&&&10\tnote{a} &\\
&&2&1600&98\\
&&&16&\\
&&4&2600&156\\
&&&26&\\
&&10&4700&287\\
&&&47&\\
\hline
\end{tabular}
\begin{tablenotes}
\item[a]{The \{1 $M_\oplus$, 0.3 AU, low $\rhog$\} model could not be evolved because its outer radiative zone was too thin to be resolved. The omission of this model is unimportant because we do not expect it
to yield a volumetrically
significant atmosphere, judging from our other 1-$M_\oplus$ models.}
\end{tablenotes}
\end{threeparttable}
\end{table}

\begin{figure}
\begin{center}
\includegraphics{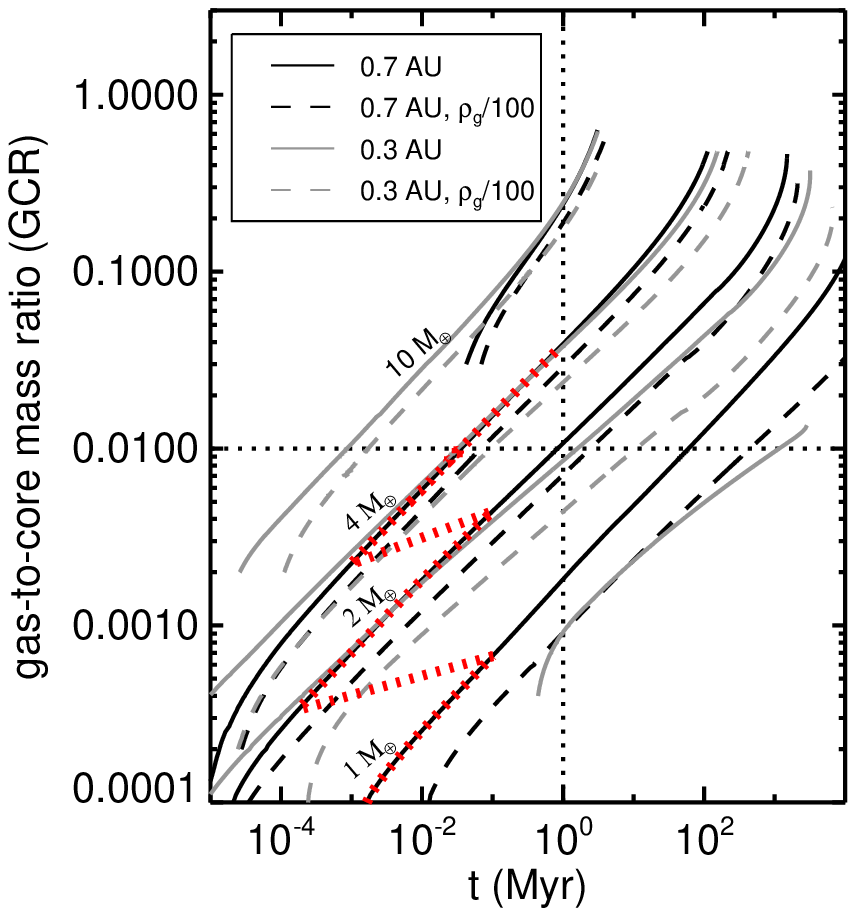}
\caption{\label{fig:atm} Gas-to-core mass fractions (GCRs) vs. time for four
core masses (labeled) at $\{a=0.3\,\au,\, T=900 \, {\rm K}\}$ (gray) and
$\{a=0.7\,\au,\, T=600 \,{\rm K}\}$ (black) with high (solid) and low (dashed)
ambient disk gas density $\rhog$ (see Table 1).
The red striped line illustrates an example ``evolutionary track"
of a body that grows from 1 to 2 to $4 M_\oplus$ through non-destructive
mergers with bodies similar in size to itself; upon each collision, we assume
that the body loses half its atmosphere and then continues accreting gas
along its new track. By the time the gas disk clears at $t = 1$ Myr (vertical
dotted line), our example 4-$M_\oplus$ core has accreted
a GCR $\approx$ 4\% atmosphere, enough to make it a voluminous
mini-Neptune.}
\end{center}
\end{figure}

Figure \ref{fig:atm} plots the resulting gas accretion histories
for dusty-atmosphere opacities. As discussed by
\citet{Lee14}, these histories are not very sensitive to
outer boundary conditions like $T$, $a$, or $\rhog$, because 
accretion is controlled by the ability of the atmosphere
to cool internally, i.e., by conditions
at the atmosphere's radiative-convective boundary which is largely
decoupled from the nebula.
From Figure \ref{fig:atm} we infer the
following rule of thumb: a core must be at least 2 $M_\oplus$ to
accrete a volumetrically significant atmosphere within 1 Myr.
(A gas depletion timescale shorter than 1 Myr
would correspond to a larger core mass for the rule of thumb.) For
example, a 2 $M_\oplus$ core located at 0.3 AU within a
disk having $\rhog = 1600 \times 10^{-12}$
g/cm$^3$ accretes a GCR $\sim$ 1\% atmosphere within 1 Myr.
If said core has an Earth-like composition, corresponding to a
radius of 1.2 $R_\oplus$, a 1\% atmosphere would inflate the
planet's total radius to 2 $R_\oplus$ \citep{Lope14}. 
For context, core masses $\gtrsim 2 M_\oplus$ also avoid complete atmospheric loss by photoevaporation at $a \sim 0.3$--1 AU,
assuming a 10\% efficiency in converting
incident XUV radiation from the host star
into the kinetic energy of a planetary outflow
(Figure 2 of \citealt{Lope13}; see also
\citealt{Owen13}).

\subsection{Combining core growth and atmospheric accretion}\label{combine}

\begin{figure}
\begin{center}
\includegraphics{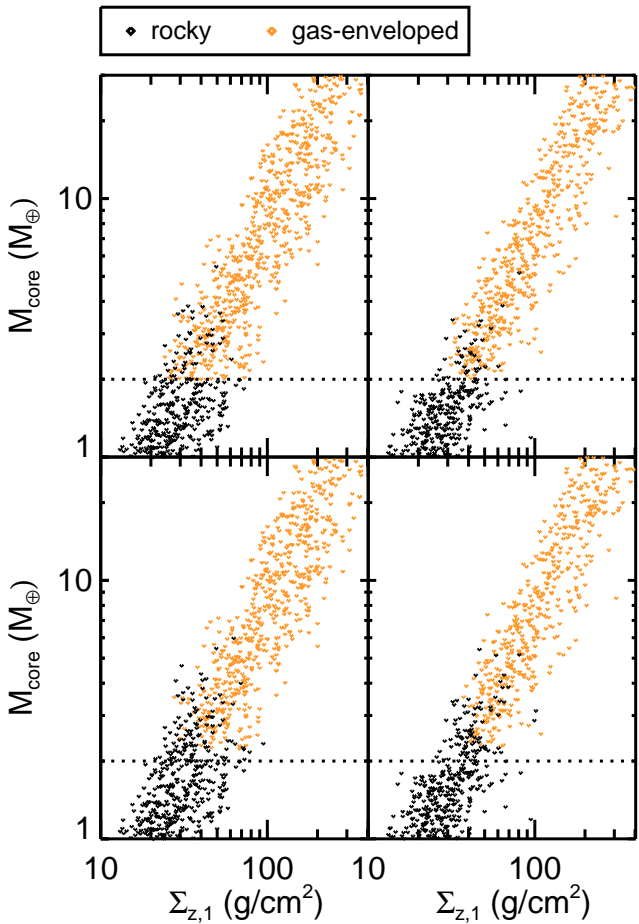}
\caption{\label{fig:comp} Compositional outcomes
and how they depend on solid surface density (as measured
by its normalization $\Sigma_{z,1}$ evaluated at 1 AU).
Top panels are based on our ``rule-of-thumb" treatment for gas
accretion: a planet is considered gas-enveloped (colored orange) if it grows to $2 M_\oplus$ in $< 1$ Myr; otherwise it stays purely rocky (black).
Bottom panels are derived from the detailed evolutionary tracks
shown in Figure \ref{fig:atm}; planets are considered gas-enveloped
if they attain a GCR $\geq$ 1\% within 1 Myr; see text for details. Left panels utilize data from the $N$-body simulations shown in Figure \ref{fig:growh} which are based on Hill scalings; right panels derive
from Figure \ref{fig:growc} which uses sound-speed scalings.
How the compositional outcome depends on solid surface density
is insensitive to modeling specifics, as judged by the similarity
of all four panels.}
\end{center}
\end{figure}

We now use the results from the atmospheric accretion simulations
(Figure \ref{fig:atm}) to identify which of the cores grown in
the $N$-body simulations (Figures \ref{fig:growh} and
\ref{fig:growc}) acquire voluminous gas envelopes vs.~stay purely
rocky. We carry out two methods of analysis.
The first and simpler procedure is to
apply the rule of thumb described in Section \ref{aam}. We
declare that all cores that grow to 2 $M_\oplus$ within 1 Myr
accrete a gas envelope, and that all cores that fail to
grow that quickly remain rocky. We plot the results of this
experiment in the top panels of Figure
\ref{fig:comp} for both our Hill-scaled and sound-speed-scaled
coagulation simulations. We see that only disks with
higher solid surface densities --- $\Sigma_{z,1} \gtrsim 50$ g/cm$^2$ --- always spawn $\gtrsim 2 M_{\oplus}$ cores
fast enough to pick up atmospheres.

The results of a second, more detailed procedure
are shown in the bottom panels of Figure \ref{fig:comp}.
Here each core is prescribed to accrete an atmosphere based on the evolutionary tracks shown in Figure \ref{fig:atm}. By interpolating these tracks, we infer a gas-to-core mass fraction, $\gcr [t,\mcore(t),\rhog(t),a(t)]$, where $\gcr [t=0]=0$.\footnote{Technically $\gcr[t=0] > 0$ because
isolation-mass embryos can accrete gas before the giant impact era begins. In practice, because the gas
accretion rate depends so steeply on core mass (see Figure \ref{fig:atm}), this initial amount of gas acquired
is negligible compared to that accreted later when the
core mass is larger.}
Let $\{t_{{\rm merger},i}\}$ be the array of merger times for a given core. For simplicity and because the dependence of the accretion rate on the gas density is weak, we assume that the gas disk (which has already been depleted to $\Sigma_{\rm gas} = \sigz$ based on the considerations in Section 2) remains in place for 1 Myr (the results are not sensitive to this exact time)
and disappears thereafter:
\begin{eqnarray}
\rhog  = & \sigz \Omega/c_{\rm s} & \,\,\,\,{\rm for}\,\,\,\, t < 1 {\rm Myr} \nonumber\\
  = & 0 & \,\,\,\,{\rm for}\,\,\,\, t \ge 1 {\rm Myr}\,. \nonumber 
\end{eqnarray}
If the first merger $t_{{\rm merger},1} > 1$ Myr, the final
atmospheric gas content of the planet equals
$\gcr [t= 1 \,{\rm Myr}, \,\mcore = \memb]$. Otherwise
we interpolate the evolutionary tracks to calculate
$\gcr[t= t_{{\rm merger},1}]$ just before
the first merger. Just after the first merger, we assume that, because of atmospheric loss from the giant impact, the new body starts with half the GCR of the larger merging body.\footnote{In practice, this atmospheric loss is not important because gas accretion occurring after the last merger prior to the disappearance of the gas disk typically dominates.} This is meant to be an approximation to the more realistic treatments developed by \citet{Schl15} and \citet{Inam15}. Note that the overwhelming majority of mergers are between bodies of comparable mass.
From there, the GCR follows a new track appropriate
to its new $M_{\rm core}$, accreting additional
gas for $t= \min (t_{{\rm merger},2}, 1 \,{\rm
Myr})-t_{{\rm merger},1}$.
This process repeats until $t_{{\rm merger},i} > 1$ Myr.
See Figure 4 for an example evolutionary trajectory.

Planets that attain GCR $\geq 1\%$ are considered gas-enveloped and colored orange in Figures \ref{fig:comp} and \ref{fig:expect}.
This second method yields essentially identical results to the
first rule-of-thumb treatment.

In all panels of Figure \ref{fig:comp}, we see that the lowest
$\sigz$ disks produce predominantly rocky (black) planets;
intermediate $\sigz$ disks yield a combination of gas-free and
gas-enveloped (orange) planets; and the highest $\sigz$ disks spawn
exclusively gas-enveloped planets. 
Figure \ref{fig:comp} indicates that virtually all planets with final masses above $\sim$5$M_\oplus$ accrete
gaseous envelopes, suggesting that radial-velocity discoveries like GJ
876 e (a 12.5 $M_\oplus$ planet with a 124-day orbital period;
\citealt{Rive10}) or 61 Vir c (a 10.6 $M_\oplus$ planet with a 38-day orbital period; \citealt{Vogt10}) are mini-Neptunes rather than super-Earths.

\citet{Inam15} also recently simulated the atmospheric mass fractions attained
by an ensemble of planets that grow from isolation mass to core mass during
the giant impact stage. Their isolation masses are formed from annuli $5 R_{\rm H}$
wide and having a solid surface density $\Sigma_z$ such that observed
exoplanets of mass $M_p$ can form from feeding zones that are
$\Delta a = 2v_{\rm esc}/\Omega$ wide, where $v_{\rm esc}$
is evaluated for $M_p$ \citep{Schl14}. They simulate the
accretion of gas during the isolation mass stage; subsequent atmospheric loss
during giant impacts; and the final post-giant impact accretion of gas. The
giant impacts are based on Monte Carlo simulations --- by comparison with the
$N$-body simulations presented here --- and atmospheric loss is treated in a
more detailed and realistic way than our simple factor-of-2 prescription.
When they consider only atmospheres accreted during the isolation-mass stage
and subsequently eroded through impacts, the resulting atmospheric masses
fractions are low, on the order of $\sim$10$^{-3}$ (their Figure 8). However, atmospheric mass
fractions reach several percent for subsequent accretion by core masses above
$2 M_\oplus$ beyond 0.15 AU (their Figure 9), unless the energy from giant
impacts happens to be released over exactly the disk dissipation timescale. Thus our results are consistent with their post-giant impact results, in which gas accretion is dominated by the largest core mass achieved before the dissipation of the gas disk.

\subsection{Effects of migration}

The formation theory we have explored above is an in-situ theory; the planets are
assumed to form in place. We sketch here how our results are impacted by orbital
migration. Transporting bodies by gravitational torques in a disk (for a review,
see \citealt{Kley12}) can: (1) supply planetary embryos
from the outer disk to the inner disk; (2) establish a resonant chain of protoplanets
in the inner disk (but see \citealt{Gold14} for reasons to believe
such resonant locks are easily broken); and (3) move fully formed super-Earths and mini-Neptunes
to the inner disk. We will argue that all these effects are compatible with
our thesis that the disk's solid surface density is a major determinant
in a planet's final gas-to-rock ratio.

In scenario (1), planet formation would proceed as
detailed above (Sections 3.1--3.3)
but with $\Sigma_z$ reflecting the surface density of
embryos supplied from the outer disk rather
than the surface density of dust/planetesimals in the
inner disk. 
The simulations of Sections 3.1--3.3 model events
post-dating the growth or delivery of embryos, so
migration scenario (1)
would not impact our conclusions qualitatively.
Scenario (2) would be characterized by larger orbital spacings between protocores
and longer orbit crossing times. Nevertheless the de-stabilization time should depend on planet
masses (see, e.g., the resonant chains exhibited by the inner moons of Uranus; \citealt{Fren15})
which in turn would reflect the isolation masses in the outer disk; thus merger
times should still depend sensitively on $\Sigma_{z,1}$. Finally, under scenario (3),
coagulation of rocky cores and accretion of gas would occur at larger orbital
distances, where isolation masses are larger and orbital timescales are longer.
The latter two effects appear to cancel for our fiducial semi-major axis range 
(0.04--1 AU); the black vs.~blue points in Figures 2 and 3 exhibit similar trends.
Whether they cancel so nearly at larger semi-major axes remains to be determined.
If they do not cancel, then the threshold $\Sigma_{z,1}$ required to produce
gas-enveloped planets could be larger or smaller, but it would still exist; i.e.,
we would still expect a population of purely rocky planets in the lowest $\Sigma_{z,1}$ disks.
Because feeding zone annuli tend to be wider at larger
orbital distances, cores formed in the outer disk 
do not need as large a
$\Sigma_{z,1}$ to produce cores that are massive enough to accrete significant amounts of gas as isolation masses.

Recent models incorporating the effects of migration have assessed the effect of a parameter
analogous to $\Sigma_{z,1}$ on the properties of super-Earths and mini-Neptunes. \citet{Coss14}
simulated planet formation with $N$-body integrations combined with prescriptive migration maps. Not treating gas accretion by cores, they found that disks with 
a higher mass in embryos delivered more massive super-Earth cores to the inner disk.
We would expect these more massive cores to more easily acquire atmospheres (Section 3.2).
In addition, \citet{Ida10} found that, for a range of migration speeds, the typical mass of super-Earths that migrate to $< 1$ AU increases with disk solid surface density normalization. The delivery of higher-mass planets to the interior regions of disks with overall higher solid density is consistent with our statement that the masses of planets that arrived by migration reflect the isolation masses in the outer disk.

\section{Comparison to the {\it Kepler} sample}

We argued above that the disk's solid surface density is critical for determining whether a planet forms rocky or gas-enveloped. Here we translate the results from Section 3 --- compositions as a function of $\sigz$ --- into observables to compare with the large \kep sample of super-Earths and mini-Neptunes. 

\subsection{Observational proxies}

\kep has discovered about 3000 planet candidates with sizes between that of Earth and Neptune. Statistical modeling of the potential astrophysical false positive population has revealed that less than 10\% of the candidates are false positives \citep{Mort11,Fres13}, allowing us to treat the sample as representative of true planets. Most \kep candidates have measured radii
but unknown masses. \citet{Lope14} recently argued that a planet's
radius can serve as a proxy for its composition (i.e., gas-to-rock
content), assuming an envelope composed of hydrogen, helium, and an
admixture of metals.\footnote{In many models, the atmospheric
metallicity can be as large as $Z \approx 0.4$.} In particular, planets larger than
$2 R_\oplus$ require a gas-to-core mass fraction $\gcr$ exceeding $\sim$1\%. This prediction has been borne out
by the \kep data. From a statistical analysis of planets with
measured masses and radii, \cite{Roge15}
(see also \citealt{Weis14} and \citealt{Wolf14})
found that the cut-off radius between planets
dense enough to be rocky and those requiring a significant volatile
component sits at $\sim$1.6$R_\oplus$; a planet of radius $1.6
R_\oplus$ has a $\sim$50\% probability of being purely rocky. 
Between 1.5 to 2 $R_\oplus$, planets appear to transition from being purely rocky
to having volumetrically significant gas envelopes (Figure 5 of \citealt{Roge15}).
A similar transition is suggested by
planets with orbital periods shorter than 1 day --- a.k.a.
ultra-short period planets or USPs. Because USPs
are so intensely irradiated by their host stars,
they may have lost their primordial atmospheres to photoevaporation
and may therefore exhibit a radius distribution corresponding
to that of purely rocky cores. All the USPs studied by
\citet{Sanc14} have radii $R < 1.68 R_\oplus$, consistent with
the interpretation that planets with larger radii possess
voluminous gaseous envelopes.

Together, these recent advances allow us to identify \kep planets
as rocky or gas-enveloped based on their radii alone. 
We pay particular attention below to \kep planets
with periods longer than $\sim$15 days, since
photoevaporation can remove the envelopes of shorter period
planets (e.g., \citealt{Lope13}),
leaving behind a planet that is purely rocky by nurture rather than
by nature.

Having opted for planet radius as a proxy for composition, we now
need a proxy for $\sigz$. Motivated by the strong correlation
between giant planet occurrence and host star metallicity
\citep{Sant01,Fisc05,Buch12,Buch14,Wang15}, we use the
spectroscopic $[M/H]$ of planet-hosting stars as a proxy
for the surface density $\sigz$ of their primordial
planet-forming disks. \kep follow-up teams have observed stellar spectra for hundreds of KOI (\kep Object of Interest) host stars. \citet{Buch14} recently published a catalogue of $[M/H]$ measured using the stellar parameter classification
(SPC) tool developed by \citet{Buch12}. Because
the host stars span a narrow stellar mass range, while $[M/H]$
spans nearly a full dex, we expect that $[M/H]$
should track $\sigz$ --- or at least its value averaged
over the entire disk (it may not track the local
value of $\sigz$ because solids can concentrate
radially in disks, so that the local disk metallicity
can deviate from the bulk stellar metallicity --- more on this
point in Section 4.3).
\citet{Buch14} reported that hosts of KOIs with 2--4 $R_\oplus$
have higher metallicities than hosts of smaller KOIs. \citet{Schla15} recently argued against the
statistical significance of this finding.

\begin{figure}
\begin{center}
\includegraphics{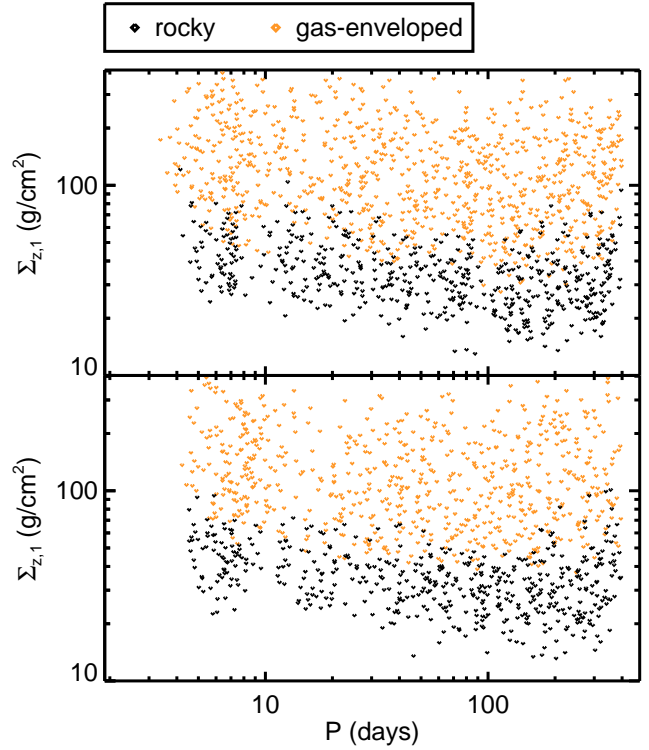}
\caption{\label{fig:expect} Compositions of planets (orange = gas-enveloped,
black = purely rocky) spawned by different solid 
surface density disks across a range of orbital periods.
Top: based on the results shown in the bottom left
panel (Hill scalings + detailed gas accretion prescription) of Figure \ref{fig:comp}.
Bottom: based on the results shown in the  bottom right panel (sound-speed
scalings + detailed gas accretion prescription) 
of Figure \ref{fig:comp}.
}
\end{center}
\end{figure}

\subsection{A metallicity divide for rocky planets}

In Figure \ref{fig:expect}, we plot the results of the
simulations from Section 3 with the added dimension of orbital
period. The transition $\sigz$ that divides
gas-enveloped planets (orange points)
from purely rocky planets (black points)
decreases slightly
with orbital period. Longer orbital periods correspond to
larger isolation masses, lowering the threshold $\sigz$ to spawn 
gas-enveloped planets only. This effect may be related to the
observed rise in the occurrence rate of Neptune-sized planets at
longer orbital periods \citep{Dong13}, though the concomitant
weakening of photoevaporation at large orbital distances 
likely also contributes. (See also
Figure 7 of \citealt{Fore14}.) Regardless, beyond 20 days, the
transition $\Sigma_{z,1}$ flattens to about 40 g/cm$^2$, about
2$\times$ the lowest $\Sigma_{z,1}$ that can spawn close-in planets
more massive than Earth, or about 4$\times$ the value of the minimum-mass solar nebula. Note that near this transition
$\Sigma_{z,1}$, a mixture of gas-enveloped and purely rocky planets
is produced.

\begin{figure}
\begin{center}
\includegraphics[width=3.5in]{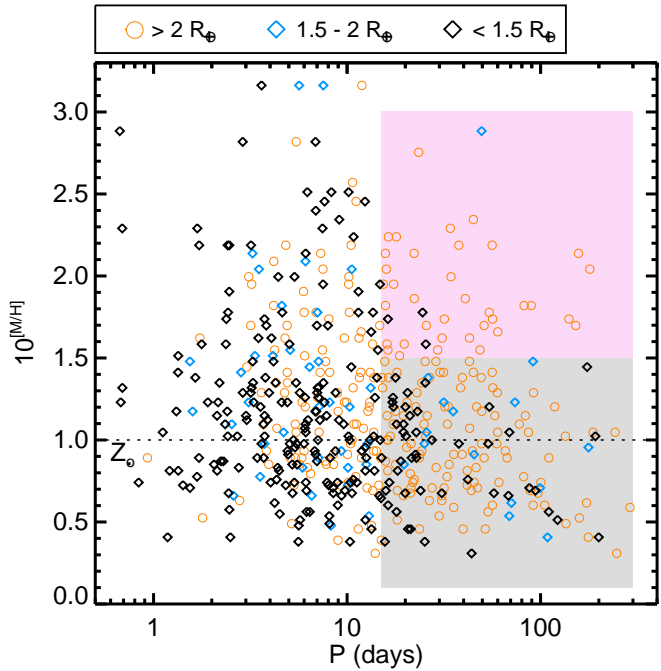}
\caption{\label{fig:obs} Metal-rich stars lack purely rocky ($R < 1.5
R_\oplus$ = black) planets on longer period orbits ($P > 15$ days); data from \citet{Buch14}. We focus on orbital periods $>$ 15 days to mitigate the influence of photoevaporation. Metal-rich stars with predominantly
gas-enveloped planets (pink box) can be distinguished
from metal-poor stars with a combination of
gas-enveloped and purely rocky planets (gray box). 
We argue that higher metallicity stars are accompanied
by disks with higher solid surface densities, which in turn
spawn $> 2 M_\oplus$ cores faster, within the gas
disk lifetime of $\sim$1 Myr; these planets more readily acquire atmospheres
and inflate their radii to $> 2R_{\oplus}$ (orange). The dotted line indicates solar metallicity for reference.
}
\end{center}
\end{figure}

Next we plot, in Figure \ref{fig:obs}, the observed host star metal
fraction (relative to solar) vs. orbital period using the
\citet{Buch14} catalogue of stars with spectroscopic parameters.
We use radius as a proxy for composition, color-coding orange those
planets with radii exceeding 2 $R_\oplus$ to signify that they
likely have volumetrically significant atmospheres. We remove KOIs
designated as false positives in the NExSci database
\citep{Mull15}. Figure \ref{fig:obs} attests that at orbital
periods $\gtrsim 15$ days, beyond the reach of
photoevaporation, metal-rich stars (in the pink box)
lack rocky planets, whereas lower metallicity stars (in the gray
box) host a mixture of rocky and gas-enveloped planets.
This trend is related to that reported by
\citet{Buch14}, who found that planets above $\sim$1.7 Earth radii
orbit higher metallicity stars. Instead of identifying
a cut in planet radius, we identify a cut in
metallicity 
above which stars host exclusively gas-enveloped, not rocky, planets.
Unlike the simulations shown in Figure 6,
the observations shown in Figure 7
show no evidence that only rocky planets orbit the lowest metallicity stars; we will
expand upon this point
in Section 4.3 (see Figure 8).

We compute a K-S statistic comparing the distributions of planetary
radii in the pink vs. gray boxes in Figure \ref{fig:obs}, yielding
a formal statistical probability of 0.00045
that the difference in the distribution of radii is due to chance.
It is not clear theoretically what the exact value of the metallicity divide $Z_{\rm div}$
should be between the pink and gray boxes;
the existence of a metallicity
divide is physically motivated by Sections 2--3 which
highlight the role played by the solid
surface density $\sigz$ in determining
a planet's final
gas content,
but the exact mapping between
$\sigz$ and $[M/H]$ is not known. Despite this
uncertainty, the high (formal) statistical confidence reflected
in the K-S test 
is robust against a range of values for $Z_{\rm div}$.
In the Appendix, we perform an alternative statistical test that
accounts for our freedom in choosing $Z_{\rm div}$; we
find there that a model with a metallicity divide is preferred to one without at 95\% confidence. The period cut of 15 days delineating the left boundaries
of the pink and gray boxes is motivated
by considerations of photoevaporation; for example,
photoevaporation has been invoked to explain
the rocky composition of Kepler-36b, which has a 13.8 day orbital period \citep{Lope13}.

In interpreting the observations, we have assumed that the reported
metallicities are correct relative to one another. Figure 7
should be remade if improved stellar parameters are obtained.
For example, if the stars reported as metal-rich were actually
giant stars, it would not have been possible to detect small
planets orbiting them, and their identification as planet-hosting
stars would be incorrect.
Another potential complication is that stars more rich in metals
tend to be larger than their metal-poor counterparts, 
making it more difficult to detect smaller (rocky) planets
orbiting the former. In the sample here, the average size of stars
that have metallicities $Z_\star > 1.5 Z_\odot$ and that host a
1--4 $R_\oplus$ KOI with a period $> 15$ days is $1.3 \pm 0.4 R_\odot$. By comparison, for stars with $Z < 1.5 Z_\odot$, the average
size is $1.1 \pm 0.4 R_\odot$.  Nevertheless, the stellar radius ranges of the two samples have substantial overlap, and the difference in the median stellar projected areas implies only a $\sim$20\% difference in signal-to-noise for transit detections. We also tried restricting the sample to stars for which otherwise identical planets of $1.3 R_\oplus$ could have been detected;
this reduced the sample size but did not change
the results qualitatively.
Therefore we interpret the trend identified in
Figure \ref{fig:obs} to be real and not just a selection effect, but
we recommend revisiting the trend as the sample size of
spectroscopically measured KOI hosts grows. We also recommend
the publication of the joint two-dimensional posteriors for stellar
metallicity and stellar radius so that any covariance in the inferred parameters can be accommodated.

\begin{figure*}
\begin{center}
\includegraphics{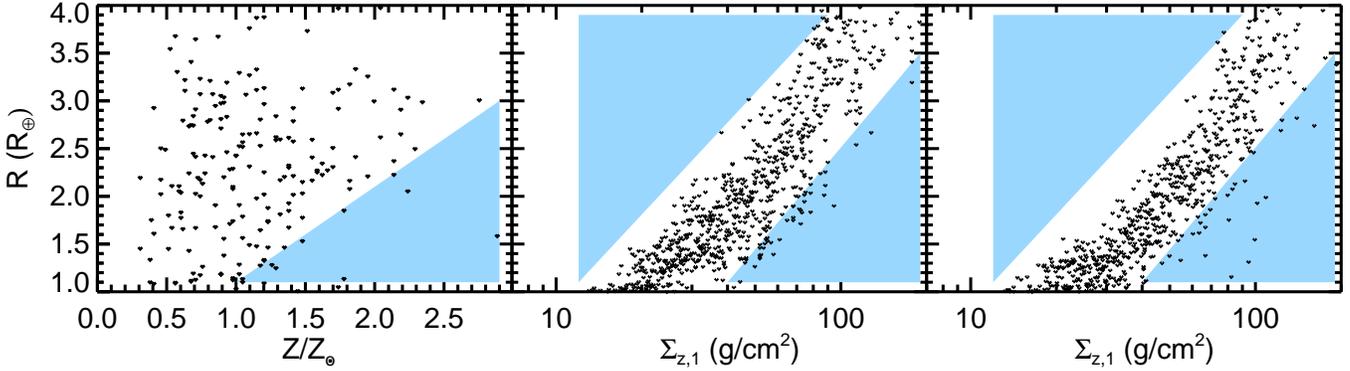}
\caption{
\label{fig:rad} Planetary radius as a function of solid surface density (its normalization $\Sigma_{z,1}$ at 1 AU) and its observational
proxy, the host star metallicity
($Z/Z_\odot$ normalized to solar). Middle and right panels are taken from
the simulations underlying the bottom panels of
Figure 5 (left and right panels, respectively), combined with the models of \citet{Lope13}.
Blue wedges are mostly empty regions of parameter space. 
Theory and observation agree on the lower right wedge --- the absence of
small radius (i.e, purely rocky) planets orbiting stars of high
metallicity (i.e., born in disks of high solid surface density). But
the upper left wedge nominally
predicted by theory is actually filled by observations
--- perhaps because radial drift of solids in the primordial disk renders
stellar metallicity an inadequate proxy of disk
solid surface density, or because
the upper left wedge is populated by planets that migrated from larger
orbital distances.
}
\end{center}
\end{figure*}

\subsection{Planet radius vs. host star metallicity in detail}

We can add another layer of modeling by computing precise
planetary radii from our simulations. The interior structure
models of \citet{Lope13} yield an empirical formula for planetary
radius as a function of core mass and GCR.
Planet radii so evaluated using the simulation data from Section 3 are plotted
against $\Sigma_{z,1}$ in Figure \ref{fig:rad}, and compared
alongside the observed distribution of planet radii vs. host star
metallicity. In both simulations and observations we see an ``empty
wedge" at large $\sigz$ (large host star metallicity) and small planetary
radius. 

However, Figure \ref{fig:rad} also reveals a discrepancy with the
observations. The simulations predict a second empty wedge: an
absence of large radius planets in the lowest $\sigz$ disks. 
But this region of parameter space is actually populated by
the observations\footnote{\citet{Schla15} found a linear correlation between planet radius and host star metallicity using a collection of candidates with sizes extending to 15 $R_\oplus$; that trend may be dominated by planets larger than 4 $R_\oplus$ so we do not compare here.}. There are a few possible explanations. First,
$[M/H]$ may be an imperfect proxy for $\Sigma_{z,1}$.
Some disks orbiting metal-poor
stars may have disproportionately large $\sigz$ if the
solids in their primordial disks drifted radially inward by
aerodynamic drag and accumulated at small orbital distances (e.g., \citealt{Youd02,Youd04,chia10,Hans12,chat14}). In other words, the
disk's solids-to-gas ratio
at any particular location can differ from the
stellar metallicity because solids and gas can segregate in disks
(see also \citealt{Andr12}).
Second, gas-enveloped planets may have formed at larger separations
beyond 1 AU --- where isolation masses are larger --- and migrated in (see Section 3.4).

\section{Conclusions}

\kep has discovered an abundance of planets with voluminous
atmospheres. A small mass percentage of volatiles ---
1\% for typical mini-Neptunes (e.g., \citealt{Wolf14}) ---
can dramatically inflate a planet's radius beyond
that of its rocky core,
with significant consequences for surface temperature, pressure,
and habitability. Gas-enveloped planets are the rule
around stars with supersolar metallicities; atmosphere-laden
planets are also found orbiting stars with subsolar metallicities,
together with purely rocky planets having practically no contribution to their
radii from their atmospheres.

What determines whether a rocky core acquires an atmosphere or not?
We have found that the protoplanetary
disk's surface density in solids is a key ingredient
in recipes for forming rocky vs. gas-enveloped planets.
In Section 2 we used order-of-magnitude scaling relations
to show that low solid surface densities prolong the timescale
for a core to grow from embryos,
so much so that by the time the core
is massive enough to acquire an atmosphere, there may not
be any disk gas left. Long coagulation timescales in low
solid surface density disks are a consequence of the larger number
of mergers required to assemble a core from smaller isolation-mass
embryos. In Section 3, we combined $N$-body
simulations of core assembly with one-dimensional 
gas accretion models to show that the highest
surface density disks produce primarily gas-enveloped planets, whereas lower solid surface
density disks can produce both gas-enveloped and rocky planets.
In Section 4, using the \citet{Buch14} spectroscopic sample,
we presented observational evidence for a lack of purely rocky
planets orbiting metal-rich stars at orbital
periods $\gtrsim$ 15 days (Figure 7 and left panel of Figure 8).
This trend is another manifestation of the correlation between
planet radius and host star metallicity for small planets reported previously
\citep{Buch14,Wang15}. If we assume a 1-to-1 
correlation between host star metallicity and disk solid surface
density, then we can reproduce the absence of purely rocky planets
around metal-rich stars (i.e., high solid surface density disks; Figure
8). But this same simple model does not reproduce the observation
that metal-poor stars also host
gas-enveloped planets with sizes up to 4 Earth radii (Figure 8).
This probably means that the assumption of a 1-to-1 correlation
between stellar metallicity and disk surface density is inadequate;
radial drift and accumulation of solids (for a review,
see \citealt{chia10}) can yield high surface density disks
(and therefore gas-enveloped planets)
even around stars with low bulk metallicity.
Alternatively, it might also be that the in-situ planet formation models used here,
which have a hard time producing ``super-puffy'' planets with
low core masses and extended atmospheres \citep[e.g.,][]{Masu14},
need to account for migration. As argued in Section 3.4,
migration should produce qualitatively
similar trends in planet composition vs.~disk solid surface density.
The subset of mini-Neptunes orbiting metal-poor stars
may have migrated from afar, where isolation masses
are larger.

These results may bear on other recently discovered trends between
host star metallicity and the properties of small planets.
 \citet[][their Figure 1]{Adib13} found that 
low-mass planets
$(\lesssim 10 M_\oplus$) orbiting metal-rich stars have orbital
periods shorter than $\sim$20 days, whereas more massive planets,
and planets orbiting metal-poor stars, span a range of orbital
periods. The restriction of mini-Neptunes/super-Earths to short
orbital periods around metal-rich stars may be a
manifestation of how embryo masses --- and presumably final
core masses --- increase with increasing orbital distance.
That is, at large orbital distances in a high surface density disk,
embryo masses may be so large that the disk cannot help
but spawn high-mass $(\gtrsim 10 M_\oplus)$ planets.
Relatedly, \citet[][their Figure 6]{Beau13} found that
super-Earths and mini-Neptunes (radii $< 4 R_\oplus$) hosted by
low-metallicity stars have orbital periods greater than 5 days;
that super-Earths and mini-Neptunes with orbital
periods shorter than 5 days are hosted by high-metallicity stars;
and that metal-poor stars do not harbor planets with radii 
$> 4 R_\oplus$. Their reported trends --- which may stem from
the higher solid surface densities expected to accompany
higher metallicity stars, in concert with the rise in embryo
mass with orbital period --- could be revisited with
the latest sample of spectroscopic metallicities from \citet{Buch14}.

We have assumed in this paper that mini-Neptunes
are either purely rocky or gas-enveloped.
The possibility that mini-Neptunes are ``water-worlds" having
a large fraction of their mass in water or ices cannot
be discounted based on mass and radius measurements
alone (see, e.g., the case of GJ 1214b; \citealt{Roge10}).
Our study favors the interpretation
that mini-Neptunes are gas-enveloped and not water-worlds;
we would not expect the prevalence of purely rocky vs.~icy
planets to be a strong function of disk solid surface density
or host star metallicity. 

Our study supports the idea that the
volatiles in mini-Neptunes are directly accreted from the
primordial nebula rather than outgassed from rock,
as outgassing does not obviously lead to the
observed trends with stellar metallicity / disk solid surface
density. Further evidence for nebular accretion can be found in
the ultra-short period ($< 1$ day) planets studied
by \citet{Sanc14}. These have a radius cut-off of
$\sim$1.7 $R_\oplus$ --- consistent with the transition from rocky
to gas-enveloped planets \citep{Roge15}, but below
the $\sim$3 $R_\oplus$ break above which the planet occurrence
rate decreases at larger orbital distances \citep{Peti13}.
The absence of planets with voluminous gas envelopes at the
shortest orbital periods is more consistent with nebular
accretion: at these close-in distances, primordially accreted atmospheres
are readily 
lost to photoevaporation during the first $\sim$0.1 Gyr of the
host star's evolution (e.g., \citealt{Owen13}), whereas
steam or gradually outgassed atmospheres can be
maintained or replenished throughout the star's lifetime.
The one notable exception of a non-rocky ultra-short period planet
is 55 Cnc e, which has an orbital period
of 0.74 days \citep{Daw10}, a radius of 2 $R_\oplus$, and a density
too low to be purely rocky \citep{Winn11}. Intriguingly,
it is situated in a system with two
close-in giant planets whose proximity to a mean-motion
resonance \citep{Marc02,McAr04,Fisc08} suggests they underwent
orbital migration. Thus 55 Cnc e 
may also have been transported from afar; it may be one
of an uncommon class of migrated icy planets, interloping
among the majority of super-Earths and mini-Neptunes 
that coagulated and acquired their atmospheres in situ.

\section*{Acknowledgments}
We are grateful to the referee for a useful
report. We thank Hilke Schlichting for several particularly helpful and
insightful discussions. We also appreciate 
valuable exchanges with Lars Buchhave, Jonathan Fortney, Brad Hansen, Jack Lissauer, Eric Lopez, 
Geoff Marcy, Sean Raymond, Leslie Rogers, Kevin Schlaufman, Scott Tremaine, and Angie Wolfgang. RID gratefully acknowledges funding by the Berkeley Miller Institute for Basic Research in Science. EC acknowledges support from grants AST-0909210 and AST-1411954 awarded by the National Science Foundation, NASA Origins grant NNX13AI57G, and Hubble Space Telescope grant HST-AR-12823.001-A.
EJL is supported in part by the Natural Sciences and Engineering Research Council of Canada under PGS D3 and the Berkeley Fellowship. Some of this research was conducted at the Kavli Institute for Theoretical Physics program on the Dynamics and Evolution of Earth-like Planets, supported in part by the National Science Foundation under Grant No. NSF PHY11-25915.

This paper makes use of the extremely valuable collection of planets discovered by the \kep Mission. Funding for the \kep mission is provided by the NASA Science Mission Directorate. We are grateful to the \kep team for all their work on this revolutionizing mission and to the ground-based follow-up teams for their host star characterization. This research has made use of the NASA Exoplanet Archive, which is operated by the California Institute of Technology, under contract with the National Aeronautics and Space Administration under the Exoplanet Exploration Program. The simulations presented here were run on the SAVIO computational cluster resource provided by the Berkeley Research Computing program at the University of California Berkeley, supported by the UC Chancellor, the UC Berkeley Vice Chancellor of Research, and the Office of the Chief Information Officer.

\bibliographystyle{mn2e}   
\bibliography{biblio.bib}

\begin{thebibliography}{86}
\expandafter\ifx\csname natexlab\endcsname\relax\def\natexlab#1{#1}\fi

\bibitem[{{Adibekyan} {et~al}\mbox{.}(2013){Adibekyan}, {Figueira}, {Santos},
  {Mortier}, {Mordasini}, {Delgado Mena}, {Sousa}, {Correia}, {Israelian}, \&
  {Oshagh}}]{Adib13}
{Adibekyan} V.~Z. {et~al.}, 2013, \aap, 560, A51

\bibitem[{{Andrews} {et~al}\mbox{.}(2012){Andrews}, {Wilner}, {Hughes}, {Qi},
  {Rosenfeld}, {{\"O}berg}, {Birnstiel}, {Espaillat}, {Cieza}, {Williams},
  {Lin}, \& {Ho}}]{Andr12}
{Andrews} S.~M. {et~al.}, 2012, \apj, 744, 162

\bibitem[{{Batalha} {et~al}\mbox{.}(2011){Batalha}, {Borucki}, {Bryson},
  {Buchhave}, {Caldwell}, {Christensen-Dalsgaard}, {Ciardi}, {Dunham},
  {Fressin}, {Gautier}, {Gilliland}, {Haas}, {Howell}, {Jenkins}, {Kjeldsen},
  {Koch}, {Latham}, {Lissauer}, {Marcy}, {Rowe}, {Sasselov}, {Seager},
  {Steffen}, {Torres}, {Basri}, {Brown}, {Charbonneau}, {Christiansen},
  {Clarke}, {Cochran}, {Dupree}, {Fabrycky}, {Fischer}, {Ford}, {Fortney},
  {Girouard}, {Holman}, {Johnson}, {Isaacson}, {Klaus}, {Machalek},
  {Moorehead}, {Morehead}, {Ragozzine}, {Tenenbaum}, {Twicken}, {Quinn},
  {VanCleve}, {Walkowicz}, {Welsh}, {Devore}, \& {Gould}}]{Bata11}
{Batalha} N.~M. {et~al.}, 2011, \apj, 729, 27

\bibitem[{{Batalha} {et~al}\mbox{.}(2013){Batalha}, {Rowe}, {Bryson},
  {Barclay}, {Burke}, {Caldwell}, {Christiansen}, {Mullally}, {Thompson},
  {Brown}, {Dupree}, {Fabrycky}, {Ford}, {Fortney}, {Gilliland}, {Isaacson},
  {Latham}, {Marcy}, {Quinn}, {Ragozzine}, {Shporer}, {Borucki}, {Ciardi},
  {Gautier}, {Haas}, {Jenkins}, {Koch}, {Lissauer}, {Rapin}, {Basri}, {Boss},
  {Buchhave}, {Carter}, {Charbonneau}, {Christensen-Dalsgaard}, {Clarke},
  {Cochran}, {Demory}, {Desert}, {Devore}, {Doyle}, {Esquerdo}, {Everett},
  {Fressin}, {Geary}, {Girouard}, {Gould}, {Hall}, {Holman}, {Howard},
  {Howell}, {Ibrahim}, {Kinemuchi}, {Kjeldsen}, {Klaus}, {Li}, {Lucas},
  {Meibom}, {Morris}, {Pr{\v s}a}, {Quintana}, {Sanderfer}, {Sasselov},
  {Seader}, {Smith}, {Steffen}, {Still}, {Stumpe}, {Tarter}, {Tenenbaum},
  {Torres}, {Twicken}, {Uddin}, {Van Cleve}, {Walkowicz}, \& {Welsh}}]{Bata13}
{Batalha} N.~M. {et~al.}, 2013, \apjs, 204, 24

\bibitem[{{Beaug{\'e}} \& {Nesvorn{\'y}}(2013)}]{Beau13}
{Beaug{\'e}} C., {Nesvorn{\'y}} D., 2013, \apj, 763, 12

\bibitem[{{Borucki} {et~al}\mbox{.}(2011{\natexlab{a}}){Borucki}, {Koch},
  {Basri}, {Batalha}, {Boss}, {Brown}, {Caldwell}, {Christensen-Dalsgaard},
  {Cochran}, {DeVore}, {Dunham}, {Dupree}, {Gautier}, {Geary}, {Gilliland},
  {Gould}, {Howell}, {Jenkins}, {Kjeldsen}, {Latham}, {Lissauer}, {Marcy},
  {Monet}, {Sasselov}, {Tarter}, {Charbonneau}, {Doyle}, {Ford}, {Fortney},
  {Holman}, {Seager}, {Steffen}, {Welsh}, {Allen}, {Bryson}, {Buchhave},
  {Chandrasekaran}, {Christiansen}, {Ciardi}, {Clarke}, {Dotson}, {Endl},
  {Fischer}, {Fressin}, {Haas}, {Horch}, {Howard}, {Isaacson}, {Kolodziejczak},
  {Li}, {MacQueen}, {Meibom}, {Prsa}, {Quintana}, {Rowe}, {Sherry},
  {Tenenbaum}, {Torres}, {Twicken}, {Van Cleve}, {Walkowicz}, \&
  {Wu}}]{Boru11a}
{Borucki} W.~J. {et~al.}, 2011{\natexlab{a}}, \apj, 728, 117

\bibitem[{{Borucki} {et~al}\mbox{.}(2011{\natexlab{b}}){Borucki}, {Koch},
  {Basri}, {Batalha}, {Brown}, {Bryson}, {Caldwell}, {Christensen-Dalsgaard},
  {Cochran}, {DeVore}, {Dunham}, {Gautier}, {Geary}, {Gilliland}, {Gould},
  {Howell}, {Jenkins}, {Latham}, {Lissauer}, {Marcy}, {Rowe}, {Sasselov},
  {Boss}, {Charbonneau}, {Ciardi}, {Doyle}, {Dupree}, {Ford}, {Fortney},
  {Holman}, {Seager}, {Steffen}, {Tarter}, {Welsh}, {Allen}, {Buchhave},
  {Christiansen}, {Clarke}, {Das}, {D{\'e}sert}, {Endl}, {Fabrycky}, {Fressin},
  {Haas}, {Horch}, {Howard}, {Isaacson}, {Kjeldsen}, {Kolodziejczak}, {Kulesa},
  {Li}, {Lucas}, {Machalek}, {McCarthy}, {MacQueen}, {Meibom}, {Miquel},
  {Prsa}, {Quinn}, {Quintana}, {Ragozzine}, {Sherry}, {Shporer}, {Tenenbaum},
  {Torres}, {Twicken}, {Van Cleve}, {Walkowicz}, {Witteborn}, \&
  {Still}}]{Boru11b}
{Borucki} W.~J. {et~al.}, 2011{\natexlab{b}}, \apj, 736, 19

\bibitem[{{Buchhave} {et~al}\mbox{.}(2014){Buchhave}, {Bizzarro}, {Latham},
  {Sasselov}, {Cochran}, {Endl}, {Isaacson}, {Juncher}, \& {Marcy}}]{Buch14}
{Buchhave} L.~A. {et~al.}, 2014, \nat, 509, 593

\bibitem[{{Buchhave} {et~al}\mbox{.}(2012){Buchhave}, {Latham}, {Johansen},
  {Bizzarro}, {Torres}, {Rowe}, {Batalha}, {Borucki}, {Brugamyer}, {Caldwell},
  {Bryson}, {Ciardi}, {Cochran}, {Endl}, {Esquerdo}, {Ford}, {Geary},
  {Gilliland}, {Hansen}, {Isaacson}, {Laird}, {Lucas}, {Marcy}, {Morse},
  {Robertson}, {Shporer}, {Stefanik}, {Still}, \& {Quinn}}]{Buch12}
{Buchhave} L.~A. {et~al.}, 2012, \nat, 486, 375

\bibitem[{{Burke} {et~al}\mbox{.}(2014){Burke}, {Bryson}, {Mullally}, {Rowe},
  {Christiansen}, {Thompson}, {Coughlin}, {Haas}, {Batalha}, {Caldwell},
  {Jenkins}, {Still}, {Barclay}, {Borucki}, {Chaplin}, {Ciardi}, {Clarke},
  {Cochran}, {Demory}, {Esquerdo}, {Gautier}, {Gilliland}, {Girouard}, {Havel},
  {Henze}, {Howell}, {Huber}, {Latham}, {Li}, {Morehead}, {Morton}, {Pepper},
  {Quintana}, {Ragozzine}, {Seader}, {Shah}, {Shporer}, {Tenenbaum}, {Twicken},
  \& {Wolfgang}}]{Burk14}
{Burke} C.~J. {et~al.}, 2014, \apjs, 210, 19

\bibitem[{{Carter} {et~al}\mbox{.}(2012){Carter}, {Agol}, {Chaplin}, {Basu},
  {Bedding}, {Buchhave}, {Christensen-Dalsgaard}, {Deck}, {Elsworth},
  {Fabrycky}, {Ford}, {Fortney}, {Hale}, {Handberg}, {Hekker}, {Holman},
  {Huber}, {Karoff}, {Kawaler}, {Kjeldsen}, {Lissauer}, {Lopez}, {Lund},
  {Lundkvist}, {Metcalfe}, {Miglio}, {Rogers}, {Stello}, {Borucki}, {Bryson},
  {Christiansen}, {Cochran}, {Geary}, {Gilliland}, {Haas}, {Hall}, {Howard},
  {Jenkins}, {Klaus}, {Koch}, {Latham}, {MacQueen}, {Sasselov}, {Steffen},
  {Twicken}, \& {Winn}}]{Cart12}
{Carter} J.~A. {et~al.}, 2012, Science, 337, 556

\bibitem[{{Chambers}(1999)}]{Cham99}
{Chambers} J.~E., 1999, \mnras, 304, 793

\bibitem[{{Chambers} \& {Wetherill}(1998)}]{Cham98}
{Chambers} J.~E., {Wetherill} G.~W., 1998, \icarus, 136, 304

\bibitem[{{Chatterjee} \& {Tan}(2014)}]{chat14}
{Chatterjee} S., {Tan} J.~C., 2014, \apj, 780, 53

\bibitem[{{Chiang} \& {Youdin}(2010)}]{chia10}
{Chiang} E., {Youdin} A.~N., 2010, Annual Review of Earth and Planetary
  Sciences, 38, 493

\bibitem[{{Cossou} {et~al}\mbox{.}(2014){Cossou}, {Raymond}, {Hersant}, \&
  {Pierens}}]{Coss14}
{Cossou} C., {Raymond} S.~N., {Hersant} F., {Pierens} A., 2014, \aap, 569, A56

\bibitem[{{Dawson} \& {Fabrycky}(2010)}]{Daw10}
{Dawson} R.~I., {Fabrycky} D.~C., 2010, \apj, 722, 937

\bibitem[{{Dong} \& {Zhu}(2013)}]{Dong13}
{Dong} S., {Zhu} Z., 2013, \apj, 778, 53

\bibitem[{{Dressing} \& {Charbonneau}(2015)}]{Dres15}
{Dressing} C.~D., {Charbonneau} D., 2015, \apj, 807, 45

\bibitem[{{Duffell}(2015)}]{Duff15}
{Duffell} P.~C., 2015, \apjl, 807, L11

\bibitem[{{Fabrycky} {et~al}\mbox{.}(2012){Fabrycky}, {Ford}, {Steffen},
  {Rowe}, {Carter}, {Moorhead}, {Batalha}, {Borucki}, {Bryson}, {Buchhave},
  {Christiansen}, {Ciardi}, {Cochran}, {Endl}, {Fanelli}, {Fischer}, {Fressin},
  {Geary}, {Haas}, {Hall}, {Holman}, {Jenkins}, {Koch}, {Latham}, {Li},
  {Lissauer}, {Lucas}, {Marcy}, {Mazeh}, {McCauliff}, {Quinn}, {Ragozzine},
  {Sasselov}, \& {Shporer}}]{Fabr12}
{Fabrycky} D.~C. {et~al.}, 2012, \apj, 750, 114

\bibitem[{{Ferguson} {et~al}\mbox{.}(2005){Ferguson}, {Alexander}, {Allard},
  {Barman}, {Bodnarik}, {Hauschildt}, {Heffner-Wong}, \& {Tamanai}}]{Ferg05}
{Ferguson} J.~W., {Alexander} D.~R., {Allard} F., {Barman} T., {Bodnarik}
  J.~G., {Hauschildt} P.~H., {Heffner-Wong} A., {Tamanai} A., 2005, \apj, 623,
  585

\bibitem[{{Fischer} {et~al}\mbox{.}(2008){Fischer}, {Marcy}, {Butler}, {Vogt},
  {Laughlin}, {Henry}, {Abouav}, {Peek}, {Wright}, {Johnson}, {McCarthy}, \&
  {Isaacson}}]{Fisc08}
{Fischer} D.~A. {et~al.}, 2008, \apj, 675, 790

\bibitem[{{Fischer} \& {Valenti}(2005)}]{Fisc05}
{Fischer} D.~A., {Valenti} J., 2005, \apj, 622, 1102

\bibitem[{{Ford} \& {Chiang}(2007)}]{Ford07}
{Ford} E.~B., {Chiang} E.~I., 2007, \apj, 661, 602

\bibitem[{{Ford} {et~al}\mbox{.}(2012{\natexlab{a}}){Ford}, {Fabrycky},
  {Steffen}, {Carter}, {Fressin}, {Holman}, {Lissauer}, {Moorhead}, {Morehead},
  {Ragozzine}, {Rowe}, {Welsh}, {Allen}, {Batalha}, {Borucki}, {Bryson},
  {Buchhave}, {Burke}, {Caldwell}, {Charbonneau}, {Clarke}, {Cochran},
  {D{\'e}sert}, {Endl}, {Everett}, {Fischer}, {Gautier}, {Gilliland},
  {Jenkins}, {Haas}, {Horch}, {Howell}, {Ibrahim}, {Isaacson}, {Koch},
  {Latham}, {Li}, {Lucas}, {MacQueen}, {Marcy}, {McCauliff}, {Mullally},
  {Quinn}, {Quintana}, {Shporer}, {Still}, {Tenenbaum}, {Thompson}, {Torres},
  {Twicken}, {Wohler}, \& {Kepler Science Team}}]{Ford12}
{Ford} E.~B. {et~al.}, 2012{\natexlab{a}}, \apj, 750, 113

\bibitem[{{Ford} {et~al}\mbox{.}(2012{\natexlab{b}}){Ford}, {Ragozzine},
  {Rowe}, {Steffen}, {Barclay}, {Batalha}, {Borucki}, {Bryson}, {Caldwell},
  {Fabrycky}, {Gautier}, {Holman}, {Ibrahim}, {Kjeldsen}, {Kinemuchi}, {Koch},
  {Lissauer}, {Still}, {Tenenbaum}, {Uddin}, \& {Welsh}}]{Ford12a}
{Ford} E.~B. {et~al.}, 2012{\natexlab{b}}, \apj, 756, 185

\bibitem[{{Ford} {et~al}\mbox{.}(2011){Ford}, {Rowe}, {Fabrycky}, {Carter},
  {Holman}, {Lissauer}, {Ragozzine}, {Steffen}, {Batalha}, {Borucki}, {Bryson},
  {Caldwell}, {Dunham}, {Gautier}, {Jenkins}, {Koch}, {Li}, {Lucas}, {Marcy},
  {McCauliff}, {Mullally}, {Quintana}, {Still}, {Tenenbaum}, {Thompson}, \&
  {Twicken}}]{Ford11}
{Ford} E.~B. {et~al.}, 2011, \apjs, 197, 2

\bibitem[{{Foreman-Mackey}, {Hogg} \& {Morton}(2014){Foreman-Mackey}, {Hogg},
  \& {Morton}}]{Fore14}
{Foreman-Mackey} D., {Hogg} D.~W., {Morton} T.~D., 2014, \apj, 795, 64

\bibitem[{{French}, {Dawson} \& {Showalter}(2015){French}, {Dawson}, \&
  {Showalter}}]{Fren15}
{French} R.~G., {Dawson} R.~I., {Showalter} M.~R., 2015, \aj, 149, 142

\bibitem[{{Fressin} {et~al}\mbox{.}(2013){Fressin}, {Torres}, {Charbonneau},
  {Bryson}, {Christiansen}, {Dressing}, {Jenkins}, {Walkowicz}, \&
  {Batalha}}]{Fres13}
{Fressin} F. {et~al.}, 2013, \apj, 766, 81

\bibitem[{{Goldreich}, {Lithwick} \& {Sari}(2004){Goldreich}, {Lithwick}, \&
  {Sari}}]{Gold04}
{Goldreich} P., {Lithwick} Y., {Sari} R., 2004, \araa, 42, 549

\bibitem[{{Goldreich} \& {Sari}(2003)}]{Gold03}
{Goldreich} P., {Sari} R., 2003, \apj, 585, 1024

\bibitem[{{Goldreich} \& {Schlichting}(2014)}]{Gold14}
{Goldreich} P., {Schlichting} H.~E., 2014, \aj, 147, 32

\bibitem[{{Hadden} \& {Lithwick}(2014)}]{Hadd14}
{Hadden} S., {Lithwick} Y., 2014, \apj, 787, 80

\bibitem[{{Hansen} \& {Murray}(2012)}]{Hans12}
{Hansen} B.~M.~S., {Murray} N., 2012, \apj, 751, 158

\bibitem[{{Hansen} \& {Murray}(2013)}]{Hans13}
{Hansen} B.~M.~S., {Murray} N., 2013, \apj, 775, 53

\bibitem[{{Hasegawa} \& {Nakazawa}(1990)}]{Hase90}
{Hasegawa} M., {Nakazawa} K., 1990, \aap, 227, 619

\bibitem[{{Ida} \& {Lin}(2010)}]{Ida10}
{Ida} S., {Lin} D.~N.~C., 2010, \apj, 719, 810

\bibitem[{{Ida} \& {Makino}(1993)}]{Ida93}
{Ida} S., {Makino} J., 1993, \icarus, 106, 210

\bibitem[{{Inamdar} \& {Schlichting}(2015)}]{Inam15}
{Inamdar} N.~K., {Schlichting} H.~E., 2015, \mnras, 448, 1751

\bibitem[{{Kley} \& {Nelson}(2012)}]{Kley12}
{Kley} W., {Nelson} R.~P., 2012, \araa, 50, 211

\bibitem[{{Kokubo} \& {Ida}(1998)}]{Koku98}
{Kokubo} E., {Ida} S., 1998, \icarus, 131, 171

\bibitem[{{Kokubo} \& {Ida}(2002)}]{Koku02}
{Kokubo} E., {Ida} S., 2002, \apj, 581, 666

\bibitem[{{Kominami} \& {Ida}(2002)}]{Komi02}
{Kominami} J., {Ida} S., 2002, \icarus, 157, 43

\bibitem[{{Lee}, {Chiang} \& {Ormel}(2014){Lee}, {Chiang}, \& {Ormel}}]{Lee14}
{Lee} E.~J., {Chiang} E., {Ormel} C.~W., 2014, \apj, 797, 95

\bibitem[{{Lissauer} {et~al}\mbox{.}(2011){Lissauer}, {Fabrycky}, {Ford},
  {Borucki}, {Fressin}, {Marcy}, {Orosz}, {Rowe}, {Torres}, {Welsh}, {Batalha},
  {Bryson}, {Buchhave}, {Caldwell}, {Carter}, {Charbonneau}, {Christiansen},
  {Cochran}, {Desert}, {Dunham}, {Fanelli}, {Fortney}, {Gautier}, {Geary},
  {Gilliland}, {Haas}, {Hall}, {Holman}, {Koch}, {Latham}, {Lopez},
  {McCauliff}, {Miller}, {Morehead}, {Quintana}, {Ragozzine}, {Sasselov},
  {Short}, \& {Steffen}}]{Liss11}
{Lissauer} J.~J. {et~al.}, 2011, \nat, 470, 53

\bibitem[{{Lissauer} {et~al}\mbox{.}(2013){Lissauer}, {Jontof-Hutter}, {Rowe},
  {Fabrycky}, {Lopez}, {Agol}, {Marcy}, {Deck}, {Fischer}, {Fortney}, {Howell},
  {Isaacson}, {Jenkins}, {Kolbl}, {Sasselov}, {Short}, \& {Welsh}}]{Liss13}
{Lissauer} J.~J. {et~al.}, 2013, \apj, 770, 131

\bibitem[{{Lithwick}, {Xie} \& {Wu}(2012){Lithwick}, {Xie}, \& {Wu}}]{Lith12}
{Lithwick} Y., {Xie} J., {Wu} Y., 2012, \apj, 761, 122

\bibitem[{{Lopez} \& {Fortney}(2013)}]{Lope13}
{Lopez} E.~D., {Fortney} J.~J., 2013, \apj, 776, 2

\bibitem[{{Lopez} \& {Fortney}(2014)}]{Lope14}
{Lopez} E.~D., {Fortney} J.~J., 2014, \apj, 792, 1

\bibitem[{{Lopez}, {Fortney} \& {Miller}(2012){Lopez}, {Fortney}, \&
  {Miller}}]{Lope12}
{Lopez} E.~D., {Fortney} J.~J., {Miller} N., 2012, \apj, 761, 59

\bibitem[{{Lubow} \& {D'Angelo}(2006)}]{Lubo06}
{Lubow} S.~H., {D'Angelo} G., 2006, \apj, 641, 526

\bibitem[{{Marcy} {et~al}\mbox{.}(2002){Marcy}, {Butler}, {Fischer},
  {Laughlin}, {Vogt}, {Henry}, \& {Pourbaix}}]{Marc02}
{Marcy} G.~W., {Butler} R.~P., {Fischer} D.~A., {Laughlin} G., {Vogt} S.~S.,
  {Henry} G.~W., {Pourbaix} D., 2002, \apj, 581, 1375

\bibitem[{{Marcy} {et~al}\mbox{.}(2014){Marcy}, {Isaacson}, {Howard}, {Rowe},
  {Jenkins}, {Bryson}, {Latham}, {Howell}, {Gautier}, {Batalha}, {Rogers},
  {Ciardi}, {Fischer}, {Gilliland}, {Kjeldsen}, {Christensen-Dalsgaard},
  {Huber}, {Chaplin}, {Basu}, {Buchhave}, {Quinn}, {Borucki}, {Koch}, {Hunter},
  {Caldwell}, {Van Cleve}, {Kolbl}, {Weiss}, {Petigura}, {Seager}, {Morton},
  {Johnson}, {Ballard}, {Burke}, {Cochran}, {Endl}, {MacQueen}, {Everett},
  {Lissauer}, {Ford}, {Torres}, {Fressin}, {Brown}, {Steffen}, {Charbonneau},
  {Basri}, {Sasselov}, {Winn}, {Sanchis-Ojeda}, {Christiansen}, {Adams},
  {Henze}, {Dupree}, {Fabrycky}, {Fortney}, {Tarter}, {Holman}, {Tenenbaum},
  {Shporer}, {Lucas}, {Welsh}, {Orosz}, {Bedding}, {Campante}, {Davies},
  {Elsworth}, {Handberg}, {Hekker}, {Karoff}, {Kawaler}, {Lund}, {Lundkvist},
  {Metcalfe}, {Miglio}, {Silva Aguirre}, {Stello}, {White}, {Boss}, {Devore},
  {Gould}, {Prsa}, {Agol}, {Barclay}, {Coughlin}, {Brugamyer}, {Mullally},
  {Quintana}, {Still}, {Thompson}, {Morrison}, {Twicken}, {D{\'e}sert},
  {Carter}, {Crepp}, {H{\'e}brard}, {Santerne}, {Moutou}, {Sobeck}, {Hudgins},
  {Haas}, {Robertson}, {Lillo-Box}, \& {Barrado}}]{Marc14}
{Marcy} G.~W. {et~al.}, 2014, \apjs, 210, 20

\bibitem[{{Masuda}(2014)}]{Masu14}
{Masuda} K., 2014, \apj, 783, 53

\bibitem[{{Mazeh} {et~al}\mbox{.}(2013){Mazeh}, {Nachmani}, {Holczer},
  {Fabrycky}, {Ford}, {Sanchis-Ojeda}, {Sokol}, {Rowe}, {Zucker}, {Agol},
  {Carter}, {Lissauer}, {Quintana}, {Ragozzine}, {Steffen}, \&
  {Welsh}}]{Maze13}
{Mazeh} T. {et~al.}, 2013, \apjs, 208, 16

\bibitem[{{McArthur} {et~al}\mbox{.}(2004){McArthur}, {Endl}, {Cochran},
  {Benedict}, {Fischer}, {Marcy}, {Butler}, {Naef}, {Mayor}, {Queloz}, {Udry},
  \& {Harrison}}]{McAr04}
{McArthur} B.~E. {et~al.}, 2004, \apjl, 614, L81

\bibitem[{{Morton} \& {Johnson}(2011)}]{Mort11}
{Morton} T.~D., {Johnson} J.~A., 2011, \apj, 738, 170

\bibitem[{{Mullally} {et~al}\mbox{.}(2015){Mullally}, {Coughlin}, {Thompson},
  {Rowe}, {Burke}, {Latham}, {Batalha}, {Bryson}, {Christiansen}, {Henze},
  {Ofir}, {Quarles}, {Shporer}, {Van Eylen}, {Van Laerhoven}, {Shah},
  {Wolfgang}, {Chaplin}, {Xie}, {Akeson}, {Argabright}, {Bachtell}, {Barclay},
  {Borucki}, {Caldwell}, {Campbell}, {Catanzarite}, {Cochran}, {Duren},
  {Fleming}, {Fraquelli}, {Girouard}, {Haas}, {He{\l}miniak}, {Howell},
  {Huber}, {Larson}, {Gautier}, {Jenkins}, {Li}, {Lissauer}, {McArthur},
  {Miller}, {Morris}, {Patil-Sabale}, {Plavchan}, {Putnam}, {Quintana},
  {Ramirez}, {Silva Aguirre}, {Seader}, {Smith}, {Steffen}, {Stewart},
  {Stober}, {Still}, {Tenenbaum}, {Troeltzsch}, {Twicken}, \&
  {Zamudio}}]{Mull15}
{Mullally} F. {et~al.}, 2015, \apjs, 217, 31

\bibitem[{{Owen} \& {Wu}(2013)}]{Owen13}
{Owen} J.~E., {Wu} Y., 2013, \apj, 775, 105

\bibitem[{{Papaloizou} \& {Larwood}(2000)}]{Papa00}
{Papaloizou} J.~C.~B., {Larwood} J.~D., 2000, \mnras, 315, 823

\bibitem[{{Petigura}, {Marcy} \& {Howard}(2013){Petigura}, {Marcy}, \&
  {Howard}}]{Peti13}
{Petigura} E.~A., {Marcy} G.~W., {Howard} A.~W., 2013, \apj, 770, 69

\bibitem[{{Rein}(2012)}]{Rein12}
{Rein} H., 2012, \mnras, 422, 3611

\bibitem[{{Rivera} {et~al}\mbox{.}(2010){Rivera}, {Laughlin}, {Butler}, {Vogt},
  {Haghighipour}, \& {Meschiari}}]{Rive10}
{Rivera} E.~J., {Laughlin} G., {Butler} R.~P., {Vogt} S.~S., {Haghighipour} N.,
  {Meschiari} S., 2010, \apj, 719, 890

\bibitem[{{Rogers}(2015)}]{Roge15}
{Rogers} L.~A., 2015, \apj, 801, 41

\bibitem[{{Rogers} {et~al}\mbox{.}(2011){Rogers}, {Bodenheimer}, {Lissauer}, \&
  {Seager}}]{Roge11}
{Rogers} L.~A., {Bodenheimer} P., {Lissauer} J.~J., {Seager} S., 2011, \apj,
  738, 59

\bibitem[{{Rogers} \& {Seager}(2010)}]{Roge10}
{Rogers} L.~A., {Seager} S., 2010, \apj, 716, 1208

\bibitem[{{Sanchis-Ojeda} {et~al}\mbox{.}(2014){Sanchis-Ojeda}, {Rappaport},
  {Winn}, {Kotson}, {Levine}, \& {El Mellah}}]{Sanc14}
{Sanchis-Ojeda} R., {Rappaport} S., {Winn} J.~N., {Kotson} M.~C., {Levine} A.,
  {El Mellah} I., 2014, \apj, 787, 47

\bibitem[{{Santos}, {Israelian} \& {Mayor}(2001){Santos}, {Israelian}, \&
  {Mayor}}]{Sant01}
{Santos} N.~C., {Israelian} G., {Mayor} M., 2001, \aap, 373, 1019

\bibitem[{{Schlaufman}(2015)}]{Schla15}
{Schlaufman} K.~C., 2015, \apjl, 799, L26

\bibitem[{{Schlichting}(2014)}]{Schl14}
{Schlichting} H.~E., 2014, \apjl, 795, L15

\bibitem[{{Schlichting}, {Sari} \& {Yalinewich}(2015){Schlichting}, {Sari}, \&
  {Yalinewich}}]{Schl15}
{Schlichting} H.~E., {Sari} R., {Yalinewich} A., 2015, \icarus, 247, 81

\bibitem[{{Steffen} {et~al}\mbox{.}(2013){Steffen}, {Fabrycky}, {Agol}, {Ford},
  {Morehead}, {Cochran}, {Lissauer}, {Adams}, {Borucki}, {Bryson}, {Caldwell},
  {Dupree}, {Jenkins}, {Robertson}, {Rowe}, {Seader}, {Thompson}, \&
  {Twicken}}]{Stef13}
{Steffen} J.~H. {et~al.}, 2013, \mnras, 428, 1077

\bibitem[{{Steffen} {et~al}\mbox{.}(2012{\natexlab{a}}){Steffen}, {Fabrycky},
  {Ford}, {Carter}, {D{\'e}sert}, {Fressin}, {Holman}, {Lissauer}, {Moorhead},
  {Rowe}, {Ragozzine}, {Welsh}, {Batalha}, {Borucki}, {Buchhave}, {Bryson},
  {Caldwell}, {Charbonneau}, {Ciardi}, {Cochran}, {Endl}, {Everett}, {Gautier},
  {Gilliland}, {Girouard}, {Jenkins}, {Horch}, {Howell}, {Isaacson}, {Klaus},
  {Koch}, {Latham}, {Li}, {Lucas}, {MacQueen}, {Marcy}, {McCauliff}, {Middour},
  {Morris}, {Mullally}, {Quinn}, {Quintana}, {Shporer}, {Still}, {Tenenbaum},
  {Thompson}, {Twicken}, \& {Van Cleve}}]{Stef12a}
{Steffen} J.~H. {et~al.}, 2012{\natexlab{a}}, \mnras, 421, 2342

\bibitem[{{Steffen} {et~al}\mbox{.}(2012{\natexlab{b}}){Steffen}, {Ford},
  {Rowe}, {Fabrycky}, {Holman}, {Welsh}, {Batalha}, {Borucki}, {Bryson},
  {Caldwell}, {Ciardi}, {Jenkins}, {Kjeldsen}, {Koch}, {Pr{\v s}a},
  {Sanderfer}, {Seader}, \& {Twicken}}]{Stef12}
{Steffen} J.~H. {et~al.}, 2012{\natexlab{b}}, \apj, 756, 186

\bibitem[{{Vogt} {et~al}\mbox{.}(2010){Vogt}, {Wittenmyer}, {Butler},
  {O'Toole}, {Henry}, {Rivera}, {Meschiari}, {Laughlin}, {Tinney}, {Jones},
  {Bailey}, {Carter}, \& {Batygin}}]{Vogt10}
{Vogt} S.~S. {et~al.}, 2010, \apj, 708, 1366

\bibitem[{{Wang} \& {Fischer}(2015)}]{Wang15}
{Wang} J., {Fischer} D.~A., 2015, \aj, 149, 14

\bibitem[{{Weiss} \& {Marcy}(2014)}]{Weis14}
{Weiss} L.~M., {Marcy} G.~W., 2014, \apjl, 783, L6

\bibitem[{{Winn} {et~al}\mbox{.}(2011){Winn}, {Matthews}, {Dawson}, {Fabrycky},
  {Holman}, {Kallinger}, {Kuschnig}, {Sasselov}, {Dragomir}, {Guenther},
  {Moffat}, {Rowe}, {Rucinski}, \& {Weiss}}]{Winn11}
{Winn} J.~N. {et~al.}, 2011, \apjl, 737, L18

\bibitem[{{Wolfgang} \& {Lopez}(2015)}]{Wolf14}
{Wolfgang} A., {Lopez} E., 2015, \apj, 806, 183

\bibitem[{{Wu} \& {Lithwick}(2013)}]{Wu13}
{Wu} Y., {Lithwick} Y., 2013, \apj, 772, 74

\bibitem[{{Xie}(2014)}]{Xie14}
{Xie} J.-W., 2014, \apjs, 210, 25

\bibitem[{{Yoshinaga}, {Kokubo} \& {Makino}(1999){Yoshinaga}, {Kokubo}, \&
  {Makino}}]{Yosh99}
{Yoshinaga} K., {Kokubo} E., {Makino} J., 1999, \icarus, 139, 328

\bibitem[{{Youdin} \& {Chiang}(2004)}]{Youd04}
{Youdin} A.~N., {Chiang} E.~I., 2004, \apj, 601, 1109

\bibitem[{{Youdin} \& {Shu}(2002)}]{Youd02}
{Youdin} A.~N., {Shu} F.~H., 2002, \apj, 580, 494

\end{thebibliography}

\appendix

\section{Statistical significance of absence of rocky planets orbiting metal-rich stars}

As an alternative statistical test that accounts for our freedom in choosing
the metallicity divide, we compute an odds ratio to compare a model with a 
metallicity divide (model 1) to one without (model 2). For the former, we
marginalize over the metallicity divide.
The odds ratio of model 1 over model 2 is:

\begin{equation}
\frac{p_{\rm model 1 | data }}{p_{\rm model 2 | data}} = \frac{p_{\rm data |  model 1}}{p_{\rm data | model 2}}\times \frac{p_{\rm model 1}}{p_{\rm model 2}} \,.
\end{equation}
If we give equal prior weight to the two models, the odds ratio becomes:

\begin{equation}
\label{eqn:odds}
\frac{p_{\rm model 1 | data }}{p_{\rm model 2 | data}} = \frac{p_{\rm data |  model 1}}{p_{\rm data | model 2}} \,.
\end{equation}

Here the data are the inferred compositions of observed planets, so we define a
probability that an observed planet is rocky, $p_{\rm rocky}$, given its
observed radius $R_{p,i}$ and uncertainty $\sigma_{Rp,i}$. The uncertainty
$\sigma_{Rp,i}$ is assumed to define a normal distribution and is computed by propagating the uncertainty from the stellar radius \citep{Buch14}
and transit depth \citep{Mull15},
assuming that the uncertainties in these quantities are independent.
For simplicity, we estimate 

\begin{eqnarray}
p_{{\rm rocky} | R_p} =& 0,& R_p > 2 R_\oplus \nonumber \\
&1,& R_p < 2 R_\oplus \,.\nonumber \\
\end{eqnarray}
We clarify that this $p_{\rm rocky}$ differs from the $p_{\rm rocky}$ defined by \citet{Roge15}, who compute $p_{\rm rocky}$ based on both mass and radius measurements. Our $p_{\rm rocky}$ abstracts the results of \citet{Roge15} and other studies \citep{Lope14,Sanc14,Weis14,Wolf14} to assert that $2 R_\oplus$ is an approximate cut-off above which a planet is very unlikely to be purely rocky. We marginalize over the uncertainty in the planet's radius:
\begin{equation}
p_{\rm rocky} = \int_0^{2 R_\oplus} \frac{1}{\sqrt{2\pi \sigma_{Rp,i}^2}} \exp\left(-\frac{(R-R_{p,i})^2}{2\sigma_{Rp,i}^2} \right) dR \,.
\end{equation}

Next we evaluate the denominator of the odds ratio in Equation \ref{eqn:odds}, the probability of the data given the model without the metallicity divide (model 2):
\begin{eqnarray}
\label{eqn:p2}
p_{\rm data | model 2} 
= \nonumber \\ \Pi_i \int_{0}^{1}\left[ f_{\rm rocky} p_{{\rm rocky}, i} + (1- f_{\rm rocky}) \left(1-  p_{{\rm rocky}, i}\right) \right] df_{\rm rocky} 
\end{eqnarray}
\noindent where $i$ refers to an individual planet.
Equation \ref{eqn:p2} marginalizes over $f_{\rm rocky}$, the fraction of the sample comprised of rocky planets (e.g., $f_{\rm rocky} = 0.5$ for an even mixture of rocky and gaseous planets).

A similar formula to Equation \ref{eqn:p2} yields
the probability of the data given
model 1 for host stars with metallicities below the divide $Z_{\rm div}$; now $f_{\rm rocky}$ is the fraction of the sample below the divide that is rocky:
\begin{eqnarray}
\label{eqn:p1b}
p_{{\rm data}_i | {\rm model 1}, Z_{\star,i} < Z_{\rm div}} =\nonumber\\ \int_{0}^{1}\left[ f_{\rm rocky} p_{{\rm rocky}, i} + (1- f_{\rm rocky}) \left(1-  p_{{\rm rocky}, i}\right) \right] df_{\rm rocky} \,.
\end{eqnarray}

By contrast, above $Z_{\rm div}$, planets are unlikely to be rocky. We marginalize over $f'_{\rm rocky}$, the fraction of the sample above the cut that is rocky, assuming $f'_{\rm rocky}$ ranges
from 0 to 10\%:
\begin{eqnarray}
\label{eqn:p1a}
p_{{\rm data}_i | {\rm model 1}, Z_{\star,i} > Z_{\rm div}} =\nonumber\\ \frac{1}{0.1}\int_{0}^{0.1}\left[ f'_{\rm rocky} p_{{\rm rocky}, i} + (1- f'_{\rm rocky}) \left(1-  p_{{\rm rocky}, i}\right) \right] df'_{\rm rocky} \,.
\end{eqnarray}
Marginalizing over $Z_{\rm div}$, we have
\begin{eqnarray}
\label{eqn:p1}
p_{\rm data | model 1}  =  \frac{1}{3} \int_{Z_{\rm div}=0}^{Z_{\rm div}=3} \nonumber \\ 
\Pi_i \left(p_{{\rm data}_i | {\rm model 1}, Z_{\star,i} < Z_{\rm div}}+p_{{\rm data}_i | {\rm model} 1, Z_{\star,i} > Z_{\rm div}}\right) dZ_{\rm div} \,.
\end{eqnarray}

We use the observed sample (Figure \ref{fig:obs}) to evaluate
Equations \ref{eqn:p2} and \ref{eqn:p1}. We thereby
obtain an odds ratio (Equation \ref{eqn:odds}) of 20 in favor of
the two-population model (i.e., 95\% confidence). The integrand in Equation \ref{eqn:p1} peaks
for $Z_{\rm div} = 1.5$.

\label{lastpage}

\end{document}